\documentclass{article}

\usepackage{graphicx}
\usepackage{amssymb}
\usepackage{amsmath}
\usepackage{mathtools}
\usepackage{float}
\usepackage[numbers]{natbib}
\usepackage{url}

\usepackage[
    left=2.5cm,
    right=2.5cm,
    top=2.5cm,
    bottom=2.5cm,
]{geometry}

\title{Theory of Dendritic Crack Growth in Ceramic Solid-State Batteries}
\author{Ansgar Lowack}
\date{March 2026}

\begin{document}

\maketitle

\begin{abstract}
\noindent In solid-state batteries, ceramic solid electrolytes are penetrated by dendrites when plating above a critical current density $J_\mathrm{crit}$. A dendrite will propagate by metal deposition at a pre-existing dendrite tip if the mechanical energy required to crack the ceramic open is less than the electrical energy (Joule heating) wasted by forcing the current to detour around the dendrite to the flat electrode surface. Based on this principle of minimal power dissipation, a dependence of $J_\mathrm{crit}\propto c_\mathrm{max}^{3/2}$ is derived in an analytical fashion. $c_\mathrm{max}$ is the length of the longest pre-existing, sufficiently thin interfacial defect. Consequentially, scattering of dendrite growth between samples must follow a Weibull-distribution, similar to the tensile strength of ceramic components but at smaller Weibull-modulus.
\end{abstract}

\section{Introduction}
Ceramic solid electrolytes are often considered promising candidates for enabling the use of lithium or sodium electrodes in next generation high energy-density batteries for room-temperature applications \cite{Janek2016, Janek2023, Randau2020}. In contrast to conventional liquid electrolytes, it is intuitive to believe that separators made from a dense ceramic solid electrolytes are a mechanical barrier which prevents dendrite penetration, following Monroe and Newman \cite{monroe2005}. However, experiments on a wide range of materials have demonstrated alkali metal penetration through ceramic solid electrolytes. 

The mechanism by which penetration happens is still under active investigation and likely not related to classical dendrites in the sense of sharp single crystals of the alkali metal. Nevertheless, the phenomenon is commonly refereed to as "dendrite penetration" in the literature, a convention which the presented study adapts. A growing body of theoretical \cite{Porz2017,Klinsmann2019, Yuan2021, mukherjee2023ingress, Esmizadeh2025, Xue2025, zhang2025atomic} and experimental \cite{Fincher2022, Athanasiou2024, Ning2023, Lowack2025, fincher2026electrochemical} work presents strong evidence that crack growth within the solid electrolyte which initiates at surface flaws is the dominant pathway for this alkali metal penetration. Assuming this hypothesis is correct, there is a notable lack of analytically derived expressions linking the critical current density in ceramic ion conductors to interfacial defect sizes and distributions. Such expressions could provide intuitive, design-oriented guidance for penetration-resistant solid electrolytes.

The present work addresses this gap by analytically deriving a formula for the critical current density in ceramic solid electrolytes, based on the semi-empirical understanding previously presented in \cite{Lowack2025}. To achieve this, interfacial defects are approximated as ellipsoids with symmetry points on the interfacial plane between electrode and solid electrolyte. This approximation turns finding the electrical potential (and hence the Joule dissipation) around a defect from only numerically solvable into a classical electrodynamics problem with known analytical solution. From Griffith's theory of crack-growth, an expression for the mechanical dissipation associated with alkali metal deposition at the tip of this defect and consequential crack propagation is calculated. Following Onsager's principle, minimizing total dissipation as the sum of Joule and mechanical dissipation yields the current density at which thermodynamics enable crack propagation. A phyically similar and more general expression has been previously numerically derived by Mukherjee et al. \cite{mukherjee2023ingress}. Following the presented model, each interfacial flaw yields a critical current density. Consequently, a weakest link behavior is postulated, where crack propagation will initiate at the largest relevant interfacial defect. As this relevant flaw size inherently scatters between samples, the critical current density itself will scatter between samples of similar preparation following a Weibull-distribution with small Weibull-modulus.

\section{Discussing the model}
\subsection{Geometrical assumptions}
\begin{figure}[H]
    \centering
    \includegraphics[width=0.5\linewidth]{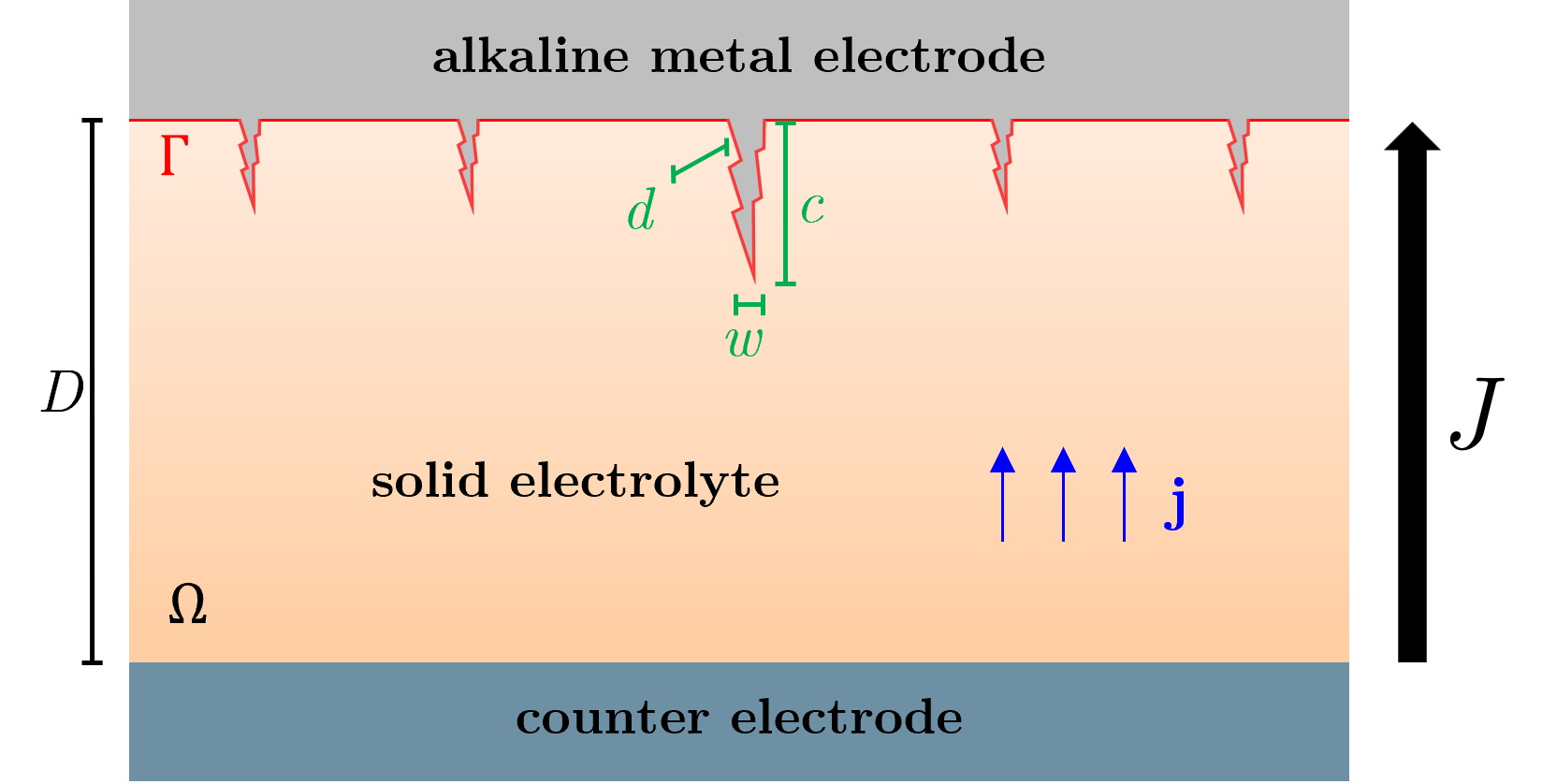}
    \caption{Schematic cross section of a SSB with unspecified counter electrode, indicating relevant variables.}
    \label{figure1}
\end{figure}
A schematic sketch of the solid-state battery is depicted in \textbf{Figure \ref{figure1}}. $\Omega$ is the volume of the solid electrolyte with thickness $D$ and cross-sectional area $A_\mathrm{cell}$. Let $\Gamma$ be the interface to the alkali metal electrode with surface area $A_\Gamma$. The solid electrolyte is assumed to be homogeneous and isotropic. 
Microstructural defects within the bulk, such as pores and grain boundaries, are neglected. 
This simplification is supported by prior experimental observations that dendrites nucleate at the interface and propagate in a stable manner once the critical current density is exceeded \cite{Lowack2025}. 
Hence, bulk defects in the solid electrolyte are not considered to play a determining role in either the nucleation mechanism or the critical current density value.

In contrast, interfacial defects act as preferential nucleation sites for dendrite formation. 
These defects are modeled as a distribution of $N$ cracks located at positions $\mathbf{x}$ along the interface $\Gamma$, described by characteristic penetration length scales $c(\mathbf{x})$, characteristic width scales $w(\mathbf{x})$ and characteristic depth scales $d$. 
Depending on the material chemistry, grain sizes, and processing conditions, $c$ is bounded below by a global minimum value $c_{\mathrm{min}}$ and is assumed to follow a Pareto distribution, yielding:
\begin{equation}
    P(C > c) = \left( \frac{c_{\min}}{c} \right)^k \quad\text{for }\quad D\gg c \geq c_{\min}, \quad k > 0
        \label{eq: Dendrit cdf-c}
\end{equation}
$k$ is the shape parameter (also frequently referred to as the Pareto index or tail index), i.e. a measure of how diverse the crack sizes are in the electrolyte ceramic: Small $k$ (e.g., $k \approx 1$) mean there is a high probability of finding very large cracks compared to the average. Large $k$ (e.g., $k > 10$) mean the cracks are very uniform in size and the material is highly consistent. 

The defects are assumed to be filled with the alkali metal, either due to alkali metal creep during electrode contacting or prior alkali metal deposition at subcritical current density.

\subsection{Transport} 
The ion transport in the solid electrolyte is described by the vector field of current density $\mathbf j(\mathbf x,J)$ which is source-free and follows from Ohm's law in $\Omega$ with ionic conductivity $\sigma$, electric field $\mathbf E(\mathbf x,J)$ and electrical potential $\Phi(\mathbf x,J)$:
\begin{equation}
\label{eq: Dendrit quellfrei und Ohm}
\nabla\cdot\mathbf j(\mathbf x, J) = 0,
\qquad
\mathbf j(\mathbf x, J) = \sigma\, \mathbf E(\mathbf x, J)=\sigma\,\nabla\Phi(\mathbf x,J),
\qquad
\mathbf x\in\Omega
\end{equation}
At the interface, $\mathbf j(\mathbf x,J)$ is subject to the global constraint:
\begin{equation}
\label{eq: Dendriten Stromdichte Gesamt}
   J=\frac{1}{A_\mathrm{cell}}\int_\Gamma j_n(\mathbf{x}, J)\,\mathrm dA,
   \qquad
   j_n(\mathbf x,J)=\mathbf j(\mathbf x,J)\cdot\mathbf n(\mathbf x),
   \qquad
   \mathbf x \in \Gamma
\end{equation}
Here $\mathbf n(\mathbf x)$ is the unit normal vector of $\Gamma$ at $\mathbf x$ and $J$ is the global current density controlled by the experimentalist. 

\subsubsection{Dissipation}
\textbf{Bulk Joule dissipation:} The current field $\mathbf{j}$ is associated with power dissipation in the solid electrolyte via Joule heating:
\begin{equation}
\label{eq:Dendriten bulk-dissipation}
\dot{Q}_\mathrm{bulk}(J)
=\int_\Omega \rho\,\big|\mathbf j(\mathbf x,J)\big|^2\,\mathrm dV,
\qquad \rho=\sigma^{-1}
\end{equation}

\textbf{Interfacial Butler-Volmer dissipation:} To deposit the alkali metal at the electrode, interfacial reaction kinetics lead to additional power dissipation, as described by the Butler-Volmer equation. This effect is neglected in this model, since it is known to be very small between alkali metal electrodes and clean oxide ceramic solid electrolytes which are sufficiently stable (and thus technologically relevant). For example Krauskopf et al. \cite{Krauskopf2020} experimentally concluded the interface resistance between lithium and LLZO to be practically $0\,\mathrm{\Omega cm^2}$ at high pressure. It is evident that this high-pressure condition is met at the tip of dendrites, which continuously propagate as cracks in the ceramic. Non-linear effects would only decrease the effect further, due to the concave relation between local interfacial overpotential and current density described by the Butler-Volmer equation.

\textbf{Interfacial mechanical dissipation}: Once the ion has crossed the interface, it must be incorporated into the lattice of the metal electrode, leading to an area-specific local volume injection rate $\dot v\big(j_n(\mathbf x,J)\big)$ with molar volume $V_{\mathrm{M}}$ of the alkali metal:
\begin{equation}
\label{eq: Dendrit Volumeninjektion}
\dot v\big(j_n(\mathbf x,J)\big) = \frac{V_{\mathrm{M}}}{F}\, j_n(\mathbf x,J)
\qquad
\mathbf x \in \Gamma
\end{equation}
Under the assumption of complete geometric contact between solid electrolyte and electrode, this volume injection dissipates mechanical power with an interfacial pressure field $p(\mathbf{x}, J)$:
\begin{equation}
\label{eq: Dendriten Mechanische Dissipation}
   \dot W(j_n(\mathbf x,J))=\int_\Gamma
\dot v\big(j_n(\mathbf x,J)\big)\,p(\mathbf{x}, J)\,\,\mathrm dA= \int_\Gamma
\frac{V_{\mathrm{M}}}{F}\,p(\mathbf{x}, J)\,j_n(\mathbf{x}, J)\,\mathrm dA
\end{equation}
Under specific circumstances this pressure field drives the dendrite formation, as discussed in the next section

\subsubsection{Interfacial pressure field}
Clearly, the value of the pressure field $p$ in equation \ref{eq: Dendriten Mechanische Dissipation} differs significantly between sites with and without defect.  

\textbf{Volume injection at site with no defect:} To deposit alkali metal at a site $\mathbf{x}_0\in\Gamma$ on the interface where there is no defect, the volume injection $\dot{V}\big(j_n(\mathbf x_0)\big)$ must be accommodated. This can be achieved by stress-driven creep in the alkali metal electrode. Generally, due to the ductile nature of lithium and sodium at room temperature, the pressure required for this is small compared to the pressure required to propagate a crack in the ceramic solid electrolyte. Thus, without any defect:
\begin{equation}
  p\big(j_n(\mathbf x_0)\big)=:p_\mathrm{creep}\big(j_n(\mathbf x_0)\big)\approx0
\end{equation}

\begin{figure}[H]
    \centering
    \includegraphics[width=0.5\linewidth]{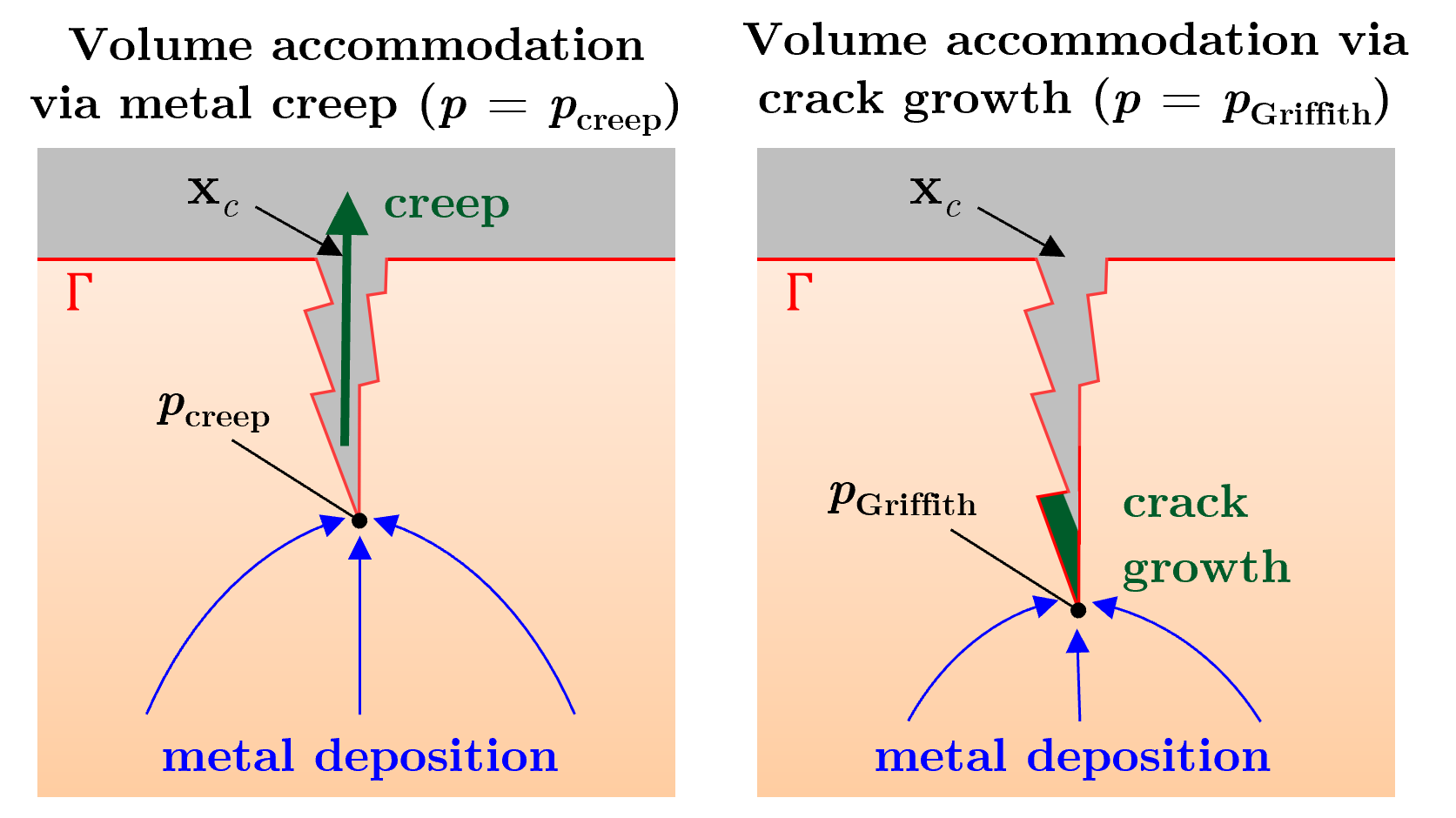}
    \caption{Schematic depiction of the alkali metal volume injection in a crack-like interfacial defect via metal creep (left) and via crack growth (right).}
    \label{figure2}
\end{figure}

\textbf{Volume injection at a defect:} The situation changes at some $\mathbf x_c\in\Gamma$, where there is a defect of length $c$, width $w$ and depth $d$. As depicted in \textbf{Figure \ref{figure2}}, now two modes of volume generation at the tip are possible.
If $\dot v(j_n(\mathbf x_c)\big)$ is small enough, again a pressure $p(j_n(\mathbf x_c)\big)=p_\mathrm{creep}(j_n(\mathbf x_c)\big)$ will build up until the creep rate in the alkali metal is large enough for volume accommodation (Figure \ref{figure2}, left). However, if the defect is mechanically confined, $p_\mathrm{creep}(j_n(\mathbf x_c)\big)$ will increase to large values. An accurate description of the metal creep in a confined micro-crack is theoretically challenging, due to the non-Newtonian behavior of alkali metal deformation under pressure and nano-scale effects. However, a rough estimation can be made by approximating the cross-section of the defect as a slit of width $w\ll d$. In this geometry, the alkali metal creep above some yield stress is approximated as viscoplastic flow in a pressure gradient $\nabla p=p/c$. Following the derivation shown e.g. by Shapovalov \cite{shapovalov2017}, this yields the total volume creep $\dot V\big(j_n(\mathbf x_c, J)\big)$ through the defect as:
\begin{equation}
    \dot V=w^2\,d\,\frac{2n}{1+2n}\bigg(\frac{p\,w}{G\,c}\bigg)^{1/n}
\end{equation}
Here, $G$ and $n$ are defined by the relation between stress $\tau$ and strain rate $\dot\gamma$ via the Norton power law:
\begin{equation}
    \tau=G\,\dot\gamma^n
\end{equation}
Normalizing $\dot V$ to the cross-section $w\cdot d$ of the slit yields:
\begin{equation}
    \label{eq: Dendriten Druck linear}
   \dot v\big(j_n(\mathbf x_c, J)\big)=\frac{\dot V\big(j_n(\mathbf x_c, J)\big)}{w\,d}=w\,\frac{2n}{1+2n}\bigg(\frac{p_\mathrm{creep}(j_n(\mathbf x_c)\big)\,w}{G\,c}\bigg)^{1/n}
\end{equation}
Rearranging and insertion of equation \ref{eq: Dendrit Volumeninjektion} yields:
\begin{equation}\label{eq: Dendriten Druck creep}
    p_\mathrm{creep}\big(j_n(\mathbf x_c, J)\big)=\bigg(\frac{V_\mathrm{M}}{F}\,\frac{1+2n}{2n}\,j_n(\mathbf x,J)\bigg)^n\frac{G\,c}{w^{1+n}}
\end{equation}
This shows that $p_\mathrm{creep}\big(j_n(\mathbf x_c, J)\big)$ grows quickly with $j_n(\mathbf x_c, J)$ for defects where $w\ll c$.

However, pressure inside such a crack cannot grow indefinitely with $j_n(\mathbf x_c, J)$. At some pressure the crack will start to propagate into the solid electrolyte, creating volume at its tip which can be filled with the depositing alkali metal. This mechanism is depicted on the right hand side in Figure \ref{figure2}. The pressure inside the defect, required to propagate the crack-like defect (opening mode I), is given by Griffith's equation: 
\begin{equation}
\label{eq: Dendrit Griffith Druck}
      p_\mathrm{Griffith}(\mathbf x_c)=\frac{K_\mathrm{Ic}}{Y_0\sqrt{c}}
\end{equation}
$K_\mathrm{Ic}$ is the fracture toughness of the solid electrolyte and $Y_0\sim 1$ is a geometrical constant which shall be omitted in the following. Notice that in the context of figure \ref{figure2}, $K_\mathrm{Ic}$ describes the fracture toughness for the alkali metal filled crack. It is smaller than the fracture toughness for a dry crack $K_\mathrm{Ic}^\mathrm{dry}$, as measured in e.g. tensile strength tests in air. Using the Griffith-Irwin-equation and Dupré's equation, the reduction follows from the surface energy $\gamma_\mathrm{ceramic}$ of the solid electrolyte, the surface energy $\gamma_\mathrm{metal}$ of the alkali metal and the interfacial energy $\gamma_\mathrm{ceramic/metal}$ as well as the work of adhesion $W_\mathrm{adh}$ between both materials:
\begin{equation}\label{eq: Dendriten KIc reduktion}
    K_\mathrm{Ic}=K_\mathrm{Ic}^\mathrm{dry}\,\sqrt{\frac{\gamma_\mathrm{ceramic/metal}}{\gamma_\mathrm{ceramic}}}=K_\mathrm{Ic}^\mathrm{dry}\,\sqrt{\frac{\gamma_\mathrm{ceramic}+\gamma_\mathrm{metal}-W_\mathrm{adh}}{\gamma_\mathrm{ceramic}}}
\end{equation}
Consider now that $p_\mathrm{creep}\big(j_n(\mathbf x_c)\big)$ can only increase with current until some value $j_\mathrm{Griffith}(\mathbf x_c)$, where:
\begin{equation}
    p_\mathrm{creep}\big(j_\mathrm{Griffith}(\mathbf x_c)\big)=p_\mathrm{Griffith}(\mathbf x_c)
\end{equation}
Insertion of equation \ref{eq: Dendriten Druck creep} and \ref{eq: Dendrit Griffith Druck} then yields:
\begin{equation}
\label{eq: Dendrit lokale krit Stromdichte}
    j_\mathrm{Griffith}(\mathbf x_c)\approx\frac{2n}{1+2n}\frac{F}{V_\mathrm{M}}\bigg(\frac{K_\mathrm{Ic}}{G\,c^{3/2}}\bigg)^{1/n}w^{(1+n)/n}
\end{equation}
This value describes the current density at or above which alkali metal must be deposited at the defect tip to thermodynamically allow crack growth. The effective area of the defect tip, where the alkali metal is deposited, scales for small $w$ like $d\cdot c$. Thus, as a necessary (but not sufficient) condition: For a dendrite to grow at $\mathbf{x}_c$, a current 
\begin{equation}
\label{eq: Dendriten Griffith Strom}
    I_c\gtrsim d\,c\,j_\mathrm{Griffith}(\mathbf x_c)=\frac{2n}{1+2n}\frac{F}{V_\mathrm{M}}\bigg(\frac{K_\mathrm{Ic}}{G}\bigg)^{1/n}w^{(1+n)/n}\,d\,c^{1-3/2n}
\end{equation}
must be deposited at the tip of a defect, to propagate the crack. $I_c$ decreases rapidly in value with decreasing $w$. Thus, only sufficiently thin defects with a low width $w$ will be dangerous in the context of dendrite formation. 

\subsection{Onsager's principle of minimal dissipation}
At a given global current density $J$, of all current fields $\mathbf j$ satisfying equation \ref{eq: Dendrit quellfrei und Ohm} to \ref{eq: Dendriten Mechanische Dissipation} mathematically, the physically realized one minimizes the total irreversible power dissipation. This variational principle is (in this specific case) a direct consequence of the second law of thermodynamics and corresponds to the colloquial term ``The electric current takes the path of lowest resistance''.

When neglecting all other minor contributions, the dissipation functional which assigns the total dissipated power to a current field $\mathbf j$ at $J$ follows from equation \ref{eq:Dendriten bulk-dissipation} and \ref{eq: Dendriten Mechanische Dissipation} as:
\begin{equation}
\label{eq: Dendriten Dissipationsfunktional}
{\mathcal D}[\mathbf j]
=
\int_\Omega \rho\, |\mathbf j(\mathbf{x}, J)|^2 \, \mathrm dV
+
\int_\Gamma
\frac{V_{\mathrm{M}}}{F}\,
p(\mathbf{x}, J)\,
j_n(\mathbf{x}, J)\,
\mathrm dA
\end{equation}

\subsection{Emergence of a local critical current density}
\begin{figure}[H]
    \centering
    \includegraphics[width=0.5\linewidth]{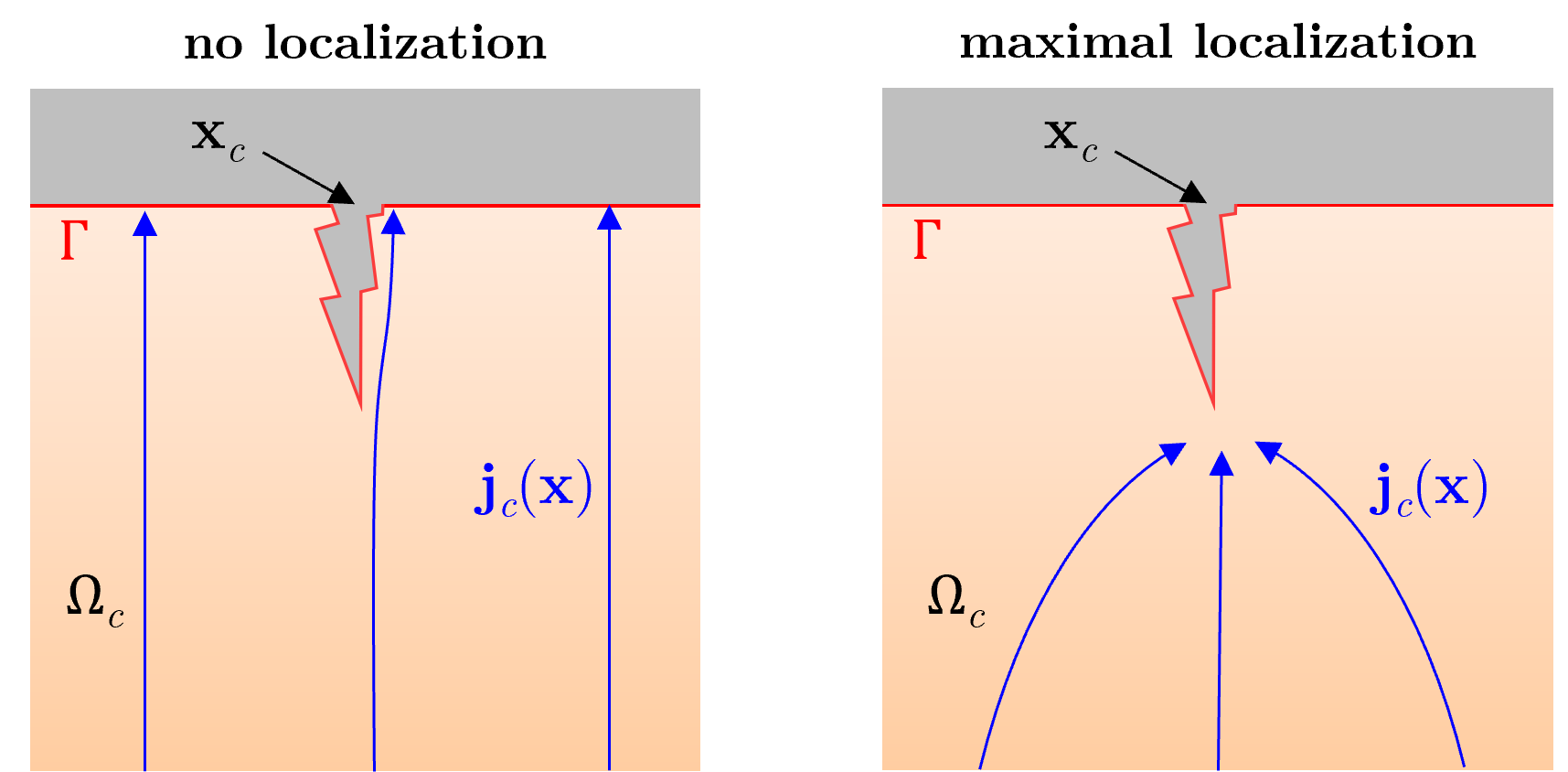}
    \caption{Schematic visualization of the current field $\mathbf j_c(\mathbf x)$ around a defect at $\mathbf x_c$ for no current localization and the maximum possible current localization.}
    \label{figure3}
\end{figure}

There is only one dendrite across the whole solid electrolyte necessary to cause an irreversible failure of the cell. The local critical current density $J^\mathrm{L}_\mathrm{crit}$ of this growing dendrite ultimately determines the observed overall critical current density. According to equation \ref{eq: Dendriten Griffith Strom} this dendrite most probably originates from a thin and well above average large interfacial defect – i.e. an 'outlier' from the high-c tail of the Pareto distribution in equation \ref{eq: Dendrit cdf-c}. Assume that these outliers, i.e. potentially dangerous defects, are spaced far enough apart to act independently from each other. This allows to analyze the failure of the cell by looking at how one single, isolated crack interacts with the current, without worrying about other cracks nearby. Thus, a correlation volume $\Omega_c\subset\Omega$ around each of the low $w/c$ ratio defects of length $c$ at $\mathbf{x}_c$. Within this region, the electrical potential is governed solely by that specific defect. Outside of this volume, the defect has no impact, allowing the electrical potential $\Phi$ and current density $\mathbf j$ to return to their uniform, bulk states:

\begin{equation}\label{eq: Dendriten Feld trivial}
    \Phi(\mathbf{x})=:
    \begin{cases}
        \Phi_c(\mathbf x) &\text{for }\mathbf x\in\Omega_c\\
        \Phi_0\approx-E_0\,z &\text{for }\mathbf x\notin\Omega_c
    \end{cases}\quad\Rightarrow
    \quad
    \mathbf j(\mathbf{x})=:
    \begin{cases}
        \mathbf j_c(\mathbf x)=-\sigma\,\nabla \Phi_c(\mathbf x) &\text{for }\mathbf x\in\Omega_c\\
        \mathbf j_0(\mathbf x)\approx\sigma\,E_0\,\mathbf{e}_z &\text{for }\mathbf x\notin\Omega_c
    \end{cases}
\end{equation}
$\sigma\,E_0=J$ is externally fixed, $z$ is the vertical component of $\mathbf{x}$ and $\mathbf{e}_z$ the unit vector in $z$-direction. This is a purely statistical assumption which requires $k$ in equation \ref{eq: Dendrit cdf-c} to be sufficiently small. For ceramic solid electrolytes this can be safely assumed as their processing is still in development - in contrast to e.g. high-end structural ceramics like $\mathrm{ZrO_2}$ where defect scattering has been reduced by decades of engineering progress. The integral constraint of equation \ref{eq: Dendriten Stromdichte Gesamt} now allows for different shapes of $\Phi_c(\mathbf x)$ and $\mathbf j_c(\mathbf x)$ in $\Omega_c$. The two extreme cases are illustrated in \textbf{Figure \ref{figure3}} and will be elaborated in the following. 

\textbf{No current localization:} If the mechanical dissipation term for metal deposition at the defect is considered infinitely large compared to the Joule dissipation term, deposition at the defect and hence current localization never minimizes the dissipation functional (equation \ref{eq: Dendriten Dissipationsfunktional}). This is always the case for sufficiently small values of $J$, since the Joule dissipation increases faster with $J$ than the mechanical dissipation. Here, $J$ is locally subcritical, $J<J^\mathrm{L}_\mathrm{crit}$. The shape of the current field is homogeneous, except for the current which must be redistributed from the volume of the field-free metal-filled defect. This is sketched in Figure \ref{figure3} on the left hand side. If the defect is sufficiently long and thin, the dissipation change due to this perturbation is small compared to the maximal localized case. Thus, the potential and current field are approximated as the trivial solution $\Phi_c(\mathbf x)=\Phi_0$ and $\mathbf j_c(\mathbf x)=\mathbf j_0$.

\textbf{Maximal current localization:} The right hand side of Figure \ref{figure3} illustrates the maximal localized current field $\mathbf j_c$ around the alkali metal filled defect - i.e. the emerging dendrite - which is \emph{electro-dynamically} allowed. The field follows from the solution $\Phi_c$ of the Laplace equation for the depicted geometry, when the whole electrolyte/metal interface (including the dendrite) is considered to be at constant electric potential as a Dirichlet boundary condition. Consequently, it represents the current field of minimal Joule dissipation. Compared to the case of no current localization, Joule dissipation is reduced by:
\begin{equation}
\label{eq: Dendriten maximale Dissipationseinspahrung}
    \Delta \dot{Q}_{\mathrm{max}}(J)=\sigma\,\int_{\Omega_c} |\nabla\Phi_c|^2\,\mathrm dV-\sigma\,\int_{\Omega_c\cup V_c} E_0^2\,\mathrm dV
\end{equation}
Here, $V_c$ is the volume inside the defect, which is filled with alkali metal. Equation \ref{eq: Dendriten maximale Dissipationseinspahrung} is the maximum potential dissipation reduction by current localization. For current localization to minimize overall dissipation and thus equation \ref{eq: Dendriten Dissipationsfunktional}, the corresponding mechanical dissipation cost $\Delta \dot{W}(J)$ cannot be larger than $\Delta \dot{Q}_{\mathrm{max}}(J)$. Since both terms in equation \ref{eq: Dendriten Dissipationsfunktional} increase with $J$ and Joule dissipation scales quadratic, while mechanical dissipation scales linearly, dendrite growth at $\mathbf x_c$ is thermodynamically possible (i.e. minimizes total dissipation in $\Omega_c$) as soon as 
\begin{equation}
\label{eq: Dendrit lokale kritische Stromdichte 3}
    \Delta \dot{Q}_{\mathrm{max}}(J=J^\mathrm{L}_\mathrm{crit})+\Delta \dot{W}(J=J^\mathrm{L}_\mathrm{crit})=0,
\end{equation} 
assuming equation \ref{eq: Dendriten Griffith Strom} is fulfilled at $\mathbf x_c$.
This defines the local critical current density $J^\mathrm{L}_\mathrm{crit}(\mathbf{x}_c)$. 

\textbf{Intuition of a local critical current density:} Consider the defect as a crack in the ceramic of characteristic width $w$, characteristic depth $d$ and characteristic length $c$ (Figure \ref{figure1}), e.g. between two grains, which comes to some sort of point at its tip. Under maximal current localization (Figure \ref{figure3}, right) it is intuitively clear that the current density will focus on this tip of the defect. This produces a volume of locally higher Joule dissipation which should scale like the volume of the defect itself, thus like $w\cdot d\cdot c$. At the foot of the defect, closer to the electrode, the current density will be decreased, resulting in a volume of reduced Joule dissipation which will also scale like $w\cdot d\cdot c$. At the base, an area scaling as $w\cdot d$ is effectively shadowed from current, as the corresponding current $I_c\propto J\,w\cdot d$ is deposited at the tip of the defect. Following equation \ref{eq: Dendriten Mechanische Dissipation}, \ref{eq: Dendrit Griffith Druck} and \ref{eq: Dendrit lokale kritische Stromdichte 3}, and considering Joule dissipation scales like $J^2$, a reasonable intuition is therefore:
\begin{equation}
\label{eq: dendriten Intuition}
    \Delta \dot{W}(J)\propto J\,p_\mathrm{Griffith}\,w\,d\propto J\frac{w\,d}{\sqrt{c}},\quad\Delta \dot{Q}_{\mathrm{max}}(J)\propto J^2\,w\,d\,c\quad \Rightarrow\quad J^\mathrm{L}_\mathrm{crit}\propto c^{-3/2}
\end{equation}
To underline this intuition mathematically for an analytically solvable example and estimate proportionality constants, the intuitively described defect is now modeled as one half of an ellipsoid. This geometry is flexible enough to believably describe a thin crack in a ceramic. Both $\Delta \dot{Q}_{\mathrm{max}}$ and $\Delta \dot{W}$ must be estimated using this approximation.

\subsection{Dissipation estimation}
\begin{figure}[H]
    \centering
    \includegraphics[width=0.5\linewidth]{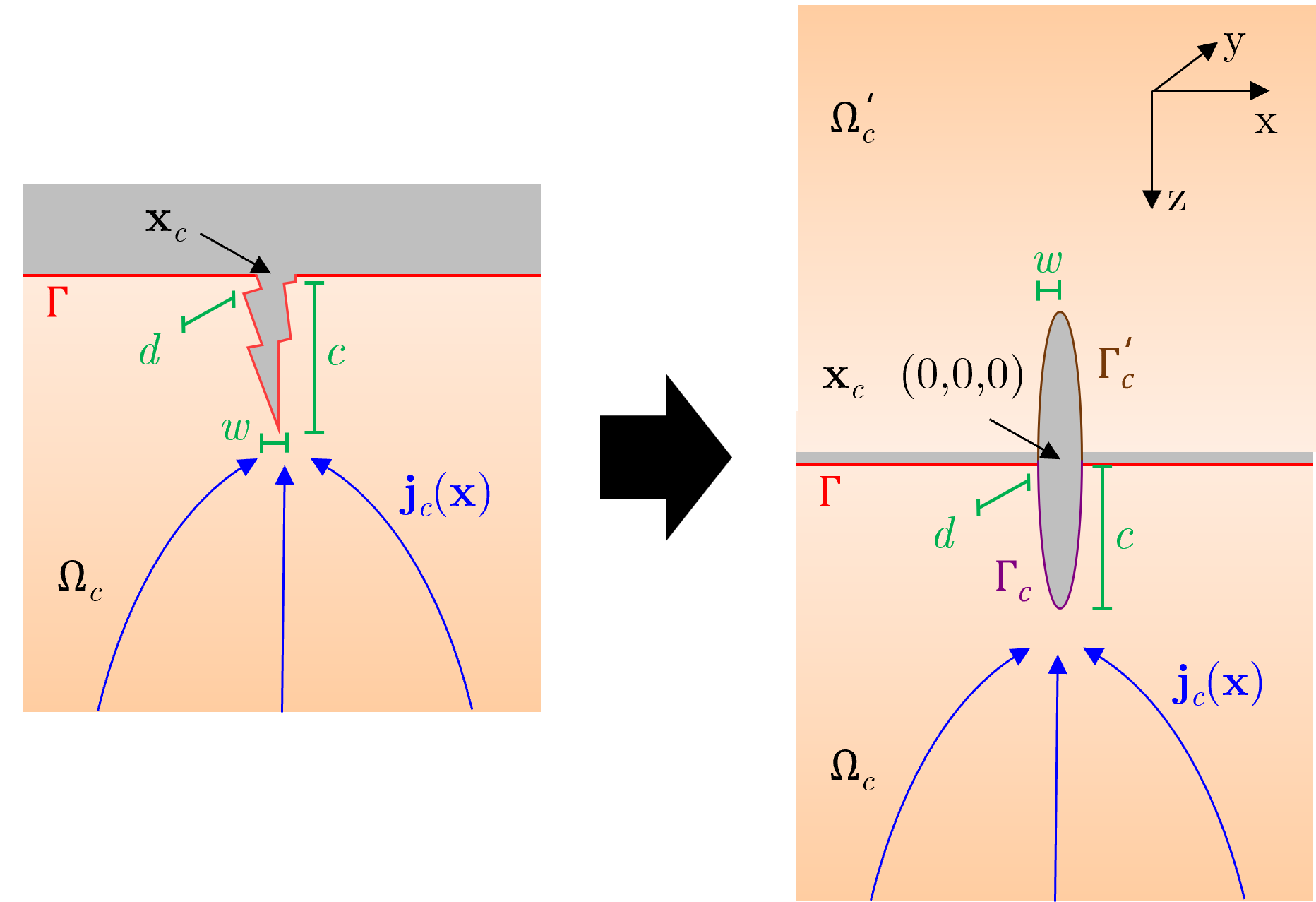}
    \caption{Approximation of defect as an ellipsoid using the method of mirror charge. The defect surface is $\Gamma_c$. Hence, the surface of the full ellipsoid is $\Gamma_c\cup\Gamma'_c$.}
    \label{figure4}
\end{figure}

\textbf{Estimating Joule dissipation}: The Laplace equation must be solved to determine $\Phi_c$ and calculate the dissipation field. For an arbitrarily shaped defect, this problem has no analytical solution. However, if the defect is approximated to be one half of an ellipsoid the problem can be treated analytically using the method of image charges. For rigorous treatment of this problem, section $3.26$ of Stratton's book "Electromagnetic Theory" \cite{stratton1941} or section $4$ of Landau \& Lifshitz's book "Electrodynamics of Continuous Media" \cite{Landau1984} is recommended. The resulting geometrical assumptions are schematically illustrated in \textbf{Figure \ref{figure4}}.

The problem is parameterized in confocal ellipsoidal coordinates $(\lambda, \mu, \nu)$ which follow from the equation
\begin{equation}\label{eq:ellcoord}
  \frac{x^2}{w^2+u} + \frac{y^2}{d^2+u} + \frac{z^2}{c^2+u} = 1
\end{equation}
as
\begin{equation}
    \lambda \in (-c^2,\infty),\quad\mu\in(-d^2,-c^2),\quad\nu\in(-w^2,d^2).
\end{equation}
In this parametrization, $\lambda=0$ describes the surface of an ellipsoid with semi-axis $w, d, c$, centered at $(0,0,0)$. This surface is defined as $\Gamma_c\cup\Gamma'_c$ in Figure \ref{figure4}. The following shorthand is defined to enhance readability: 
\begin{equation}\label{eq: R(s)}
    R(s) = \sqrt{(w^2+s)(d^2+s)(c^2+s)}
\end{equation}

Following Stratton (\S\,3.26, equation (26) in \cite{stratton1941}) or Landau \& Lifshitz (\S\,4, equation (4.24) in \cite{Landau1984}), the exterior electrostatic potential of a conducting ellipsoid held at zero potential in a uniform field $E_0\,\mathbf e_z$ is
\begin{equation}\label{eq: Stratton}
  \Phi(\lambda,z) = -E_0\,z\!\left[1 - \frac{\mathcal I(\lambda)}{\mathcal I(0)}\right],
\end{equation}
with the ellipsoidal integral:
\begin{equation}\label{eq:Ic}
  \mathcal I(\lambda) =\frac{w\,d\,c}{2}\int_\lambda^{\infty}
    \frac{ \mathrm ds}{(s + c^2)\,R(s)}
\end{equation}

The interface between electrode and solid electrolyte is introduced as a mirror plane through the ellipsoid at $z=0$. By this method of image charges, the potential given by Stratton (equation \ref{eq: Stratton}) is the potential $\Phi(\lambda,z>0)=\Phi_c(\lambda,z)$ in the volume $\Omega_c$ inside the solid electrolyte around the ellipsoid, defined only on the lower half-space $z > 0$. Consequently, the surface of the defect is $\Gamma_c$ in Figure \ref{figure4}.

Equation \ref{eq: Stratton} is now inserted into equation \ref{eq: Dendriten maximale Dissipationseinspahrung}. This yields after some calculation (shown in section 
\ref{Herleitung Joule Delta}):
\begin{equation}
\label{eq: Delta Joule Dissipation}
    \Delta \dot{Q}_{\mathrm{max}}(J)
    \approx -\frac{2\pi\,\sigma\,E_0^2\,w\,d\,c}{3\,\mathcal I(0)}
    = -\frac{2\pi\,J^2\,w\,d\,c}{3\,\sigma\,\mathcal I(0)}
\end{equation}
Thus confirming the intuition (equation \ref{eq: dendriten Intuition}). The geometric shape parameter enters through $\mathcal I(0)$ (equation \ref{eq:Ic}) which has no closed-form expression. 

\textbf{Mechanical dissipation}: The total current $I_c$ which must be deposited at the defect to realize the reduction of Joule dissipation in equation \ref{eq: Dendriten maximale Dissipationseinspahrung} is
\begin{equation}
\label{eq: Dendriten Absolutstrom}
    I_c=\int_{\Gamma_c}\mathbf j_c\cdot\mathbf n\,\mathrm{d}A=-\sigma\int_{\Gamma_c}\nabla\Phi_c\cdot\mathbf n\,\mathrm{d}A,
\end{equation}
where $\mathbf{n}$ is the outward unit normal on the ellipsoid surface $\Gamma_c$. The computation of this integral after 
insertion of the ellipsoid potential (equation \ref{eq: Stratton}), with the calculation shown in section \ref{Herleitung I}, yields:
\begin{equation}
\label{eq: Absolutstrom 2}
    I_c \approx \frac{\pi\,\sigma\,E_0\,w\,d}{\mathcal I(0)}
    = \frac{\pi\,J\,w\,d}{\mathcal I(0)}
\end{equation}

If the defect is sufficiently thin, $I_c$ will always fulfill equation \ref{eq: Dendriten Griffith Strom} and thus the mechanical dissipation change associated with current localization follows from equation \ref{eq: Dendriten Mechanische Dissipation} and \ref{eq: Dendrit Griffith Druck} as:
\begin{equation}
\label{eq: Dendrit Delta Mechanical Dissipation}
    \Delta \dot W(J)=\frac{V_\mathrm M}{F}\,p_\mathrm{Griffith}(\mathbf x_c)\,I_c\approx\frac{\pi\,V_\mathrm M\,K_\mathrm{Ic}\,J}{F\,\mathcal I(0)}\frac{w\,d}{\sqrt{c}}
\end{equation}
Again, the intuition (equation \ref{eq: dendriten Intuition}) is confirmed. The geometric shape parameter enters through $\mathcal I(0)$ (equation \ref{eq:Ic}). 

\subsection{Global critical current density}
The estimations in equation \ref{eq: Delta Joule Dissipation} and \ref{eq: Dendrit Delta Mechanical Dissipation} can now be inserted into the local critical current criterion of equation \ref{eq: Dendrit lokale kritische Stromdichte 3}. The integral $\mathcal I(0)$ cancels out and rearranging yields the local critical current density:
\begin{equation}\label{eq: Dendriten lokale kritische Stromdichte}
    J^\mathrm{L}_\mathrm{crit}(\mathbf x_c)\approx\frac{3\,\sigma\,V_\mathrm{M}\,K_\mathrm{Ic}}{2\,F\,c^{3/2}}\quad\Rightarrow\quad J^\mathrm{L}_\mathrm{crit}\approx\frac{Y\,\sigma\,V_\mathrm{M}\,K_\mathrm{Ic}}{F\,c^{3/2}}
\end{equation}
In this derivation, the geometry constant of the intuitive derivation in equation \ref{eq: dendriten Intuition} is $3/2$. Reasonably, for differently shaped defects which are not perfectly describable as an ellipsoid, the functional dependence should remain approximately the same, but the numerical prefactor might deviate. Hence, it is substituted by a geometric constant $Y\sim3/2$. As it turns out, equation \ref{eq: Dendriten lokale kritische Stromdichte} follows as a limit of the more advanced partially numerical model of Mukherjee et al. with different boundary conditions \cite{mukherjee2023ingress}.

Equation \ref{eq: Dendriten lokale kritische Stromdichte} applies for all $\mathbf{x}_c\in\Gamma$ with low width $w$ defects of length $c$. The weakest link of $\Gamma$ - i.e. the dendrite which will first penetrate the solid electrolyte - is the defect with the longest length $c=c_\mathrm{max}$. The global critical current density thus follows as:
\begin{equation}
\label{eq: Dendriten Globale kritische Stromdichte}
    \boxed{
    J_\mathrm{crit}\approx\frac{Y\,\sigma\,V_\mathrm{M}\,K_\mathrm{Ic}}{F\,c_\mathrm{max}^{3/2}}
    }
\end{equation}
In the rough approximation of viscoplastic flow, the term "sufficiently thin" can be specified by insertion of equation \ref{eq: Absolutstrom 2} in equation \ref{eq: Dendriten Griffith Strom}. The defect of length $c_\mathrm{max}$ in equation \ref{eq: Dendriten Globale kritische Stromdichte} must have a sufficiently small width $w$ to fulfill:
\begin{equation}
\label{eq: Dendriten kritische Stromdichte Notwendige Bedinung}
    J_\mathrm{crit}\gtrsim \frac{2n}{1+2n}\frac{F\,\mathcal I(0)}{\pi\,V_\mathrm{M}}\bigg(\frac{K_\mathrm{Ic}}{G}\bigg)^{1/n}c_\mathrm{max}^{1-3/2n}\,w^{1/n}
\end{equation}
Here, the dependence on $\mathcal I(0)$ remains. As Landau \& Lifshitz demonstrate (\S\,4, equation (4.29) in \cite{Landau1984}) $\mathcal I(0)\leq1$, depending on the shape of the ellipsoid (and independent of its volume). For $w=d\ll c$:
\begin{equation}\label{eq: Dendrit prolate I0}
    \mathcal I(0)=\frac{w^2}{c^2}\ln\bigg(\frac{c}{w}\bigg)
\end{equation}
\subsection{Stability and propagation}
Equation \ref{eq: Dendriten Globale kritische Stromdichte} describes the initiation of crack propagation for $J\geq J_\mathrm{crit}$ at $\mathbf x_{c_\mathrm{max}}$. Once the crack propagates, volume is generated at the crack tip and the alkali metal can plate there again without dissipating mechanical power. Hence the pressure $p(\mathbf x_{c_\mathrm{max}})$ (equation \ref{eq: Dendrit Griffith Druck}) inside the crack drops rapidly with increasing $c$, for sure faster than $c^{-1/2}$. Therefore the stress intensity $K_\mathrm{I}$ around the crack becomes subcritical and the crack stops propagating almost instantly:
\begin{equation}
    K_\mathrm{I}(\mathbf x_{c_\mathrm{max}})=Y_0\,p(\mathbf x_{c_\mathrm{max}})\,\sqrt{c_\mathrm{max}}\rightarrow0\quad \Rightarrow \quad K_\mathrm{I}(\mathbf x_{c_\mathrm{max}})<K_\mathrm{Ic}
\end{equation}
However, if $J$ is kept constant, the newly generated volume will fill up again and $J\geq J_\mathrm{crit}$ is again reached, easier this time as $c_\mathrm{max}$ has grown. Thus, once it starts propagating, the crack is inherently unstable at constant $J$ and its growth rate increases with $J$. However, it will be filled with the alkali metal as experimentally observed in \cite{Lowack2025}.

\subsection*{Weibull-statistics of critical current density scattering}
Equation \ref{eq: Dendriten Globale kritische Stromdichte} demonstrates a dependence of $J_\mathrm{crit}$ on the largest defect $c_\mathrm{max}$ at the interface - provided the defect is sufficiently thin to constrain alkali metal creep inside it (equation \ref{eq: Dendriten kritische Stromdichte Notwendige Bedinung}). In 1939 Weibull applied the same weakest link statistics argument to Griffith’s equation for mechanical strength to describe scattering between ceramic samples. Weibull’s approach can be extended to dendrite growth statistics, as shown in the following.

\textbf{Largest defect in a sample:} Assuming the $N$ defects on $\Gamma$ are statistically independent, their Pareto distribution (equation \ref{eq: Dendrit cdf-c}) implies the cumulative distribution function of the largest defect in a given sample:
\begin{equation}
    P(c_\mathrm{max}< c)=\big[P(C< c)\big]^N=\bigg[1-\bigg(\frac{c_\mathrm{min}}{c}\bigg)^k\,\bigg]^N\qquad\text{for}\qquad c\geq c_\mathrm{min}.
\end{equation}

\textbf{Cumulative distribution function of $j_\mathrm{crit}$:} For readability, the constant $\kappa$ is introduced to rewrite equation \ref{eq: Dendriten Globale kritische Stromdichte} as:
\begin{equation}
    J_\mathrm{crit}=:\kappa\,c_\mathrm{max}^{-3/2}
\end{equation}
Since $J_\mathrm{crit}$ decreases as $c_\mathrm{max}$ increases, there exists some value $\tilde{c}$ such that:
\begin{equation}
    J_\mathrm{crit}\leq \tilde{J}_\mathrm{crit}:=\kappa\,\tilde{c}^{-3/2}\quad \Leftrightarrow\quad \tilde{c}=\bigg(\frac{\kappa}{\tilde{J}_\mathrm{crit}}\bigg)^{2/3}\leq c_\mathrm{max}
\end{equation}
Consequently, the scattering of $J_\mathrm{crit}$ can be expressed by the scattering of $\tilde{c}$ as:
\begin{equation}
\label{eq: Dendriten CDF J_crit}
    P(J_\mathrm{crit}\leq \tilde{J}_\mathrm{crit})=P(c_\mathrm{max}\geq \tilde{c})=1-P(c_\mathrm{max}<\tilde{c})=1-\bigg[1-\bigg(\frac{c_\mathrm{min}}{\tilde{c}}\bigg)^k\,\bigg]^N
\end{equation}

\textbf{lower tail approximation:} The technologically relevant case of determining the statistics for $\tilde{c}\gg c_\mathrm{min}$ yields $\big(c_\mathrm{min}/\tilde{c})^k\ll 1$. Therefore, and due to $N\gg 1$ equation \ref{eq: Dendriten CDF J_crit} can be approximated as:
\begin{equation}
    P(J_\mathrm{crit}\leq \tilde{J}_\mathrm{crit})=1-\bigg[1-\bigg(\frac{c_\mathrm{min}}{\tilde{c}}\bigg)^k\,\bigg]^N\approx1-\exp\bigg[-N\,\bigg(\frac{c_\mathrm{min}}{\tilde{c}}\bigg)^k\,\bigg]
\end{equation}
Partitioning the interfacial area $A_\Gamma$ into $N$ average correlation areas $A_c$ per relevant interfacial defect, and reinsertion of $\tilde{c}$ and $\kappa$, yields:
\begin{equation}
    P(J_\mathrm{crit}\leq \tilde{J}_\mathrm{crit})\approx1-\exp\bigg[-\frac{A_\Gamma}{A_c}\,\bigg(\frac{F\,c_\mathrm{min}^{3/2}\,\tilde{J}_\mathrm{crit}}{Y\,\sigma\,V_\mathrm{M}\,K_\mathrm{Ic}}\bigg)^{2k/3}\,\bigg]
\end{equation}
With Weibull-modulus $m$ and scale parameter $J_0$, where
\begin{equation}
\label{weibull modul}
    m=\frac{2k}{3}\qquad \text{and}\qquad J_0=\frac{Y\,V_\mathrm{M}\,K_\mathrm{Ic}\,\sigma}{F\,c_\mathrm{min}^{3/2}},
\end{equation}
this is a Weibull-distribution:
\begin{equation}
    \boxed{P(J_\mathrm{crit}\leq \tilde{J}_\mathrm{crit})\approx1-\exp\bigg[-\frac{A_\Gamma}{A_c}\bigg(\frac{\tilde{J}_\mathrm{crit}}{J_0}\bigg)^m\,\bigg]}
\end{equation}
In a similar manner, a shifted Weibull-distribution can be derived, by fixing an upper cutoff $C_\mathrm{max}$ for the Pareto distribution of defect lengths in equation \ref{eq: Dendrit cdf-c}, i.e. a largest allowable defect size $C_\mathrm{max}$ that can be present in any sample. This then shifts the Weibull-distribution by a threshold
\begin{equation}
    J_u=\frac{Y\,\sigma\,V_\mathrm{M}\,K_\mathrm{Ic}}{F\,C_\mathrm{max}^{3/2}}
\end{equation}
below which no dendrites can be observed in any sample:
\begin{equation}
    \boxed{P(J_\mathrm{crit}\leq \tilde{J}_\mathrm{crit})\approx1-\exp\bigg[-\frac{A_\Gamma}{A_c}\bigg(\frac{\tilde{J}_\mathrm{crit}-J_u}{J_0}\bigg)^m\,\bigg]}
\end{equation}

\subsection{Plausibility discussion}
\textbf{Direct evidence:} Athanasiou, Fincher et al. have directly measured the stress intensity $K_\mathrm{I}$ at resting and at propagating surface flaws filled with lithium \cite{Athanasiou2024, Fincher2022}. Their values fall into the range of the fracture toughness $K_\mathrm{Ic}\sim0.5-2\,\mathrm{MPa\,\sqrt{m}}$ expected for oxide ceramics. This supports the central assumption of the model, namely that penetration proceeds by mechanical crack growth from interfacial defects once $K_\mathrm{I}\gtrsim K_\mathrm{Ic}$.

Subcritical crack growth, that is propagation at $K_\mathrm{I}<K_\mathrm{Ic}$ but above some stress intensity threshold $K_\mathrm{Iscc}$ due to tensile stress and some chemical reaction, would complicate this picture. Fincher et al. \cite{fincher2026electrochemical} report such a contribution in the form of electrochemical corrosion accompanying dendrite growth. However, penetration behaves very similarly across chemically very different solid electrolytes, while no subcritical mechanism has so far been identified which is independent of the electrolyte chemistry. Subcritical contributions are therefore unlikely to control the critical current density, even where they are present.

\begin{figure}
    \centering
    \includegraphics[width=0.9\linewidth]{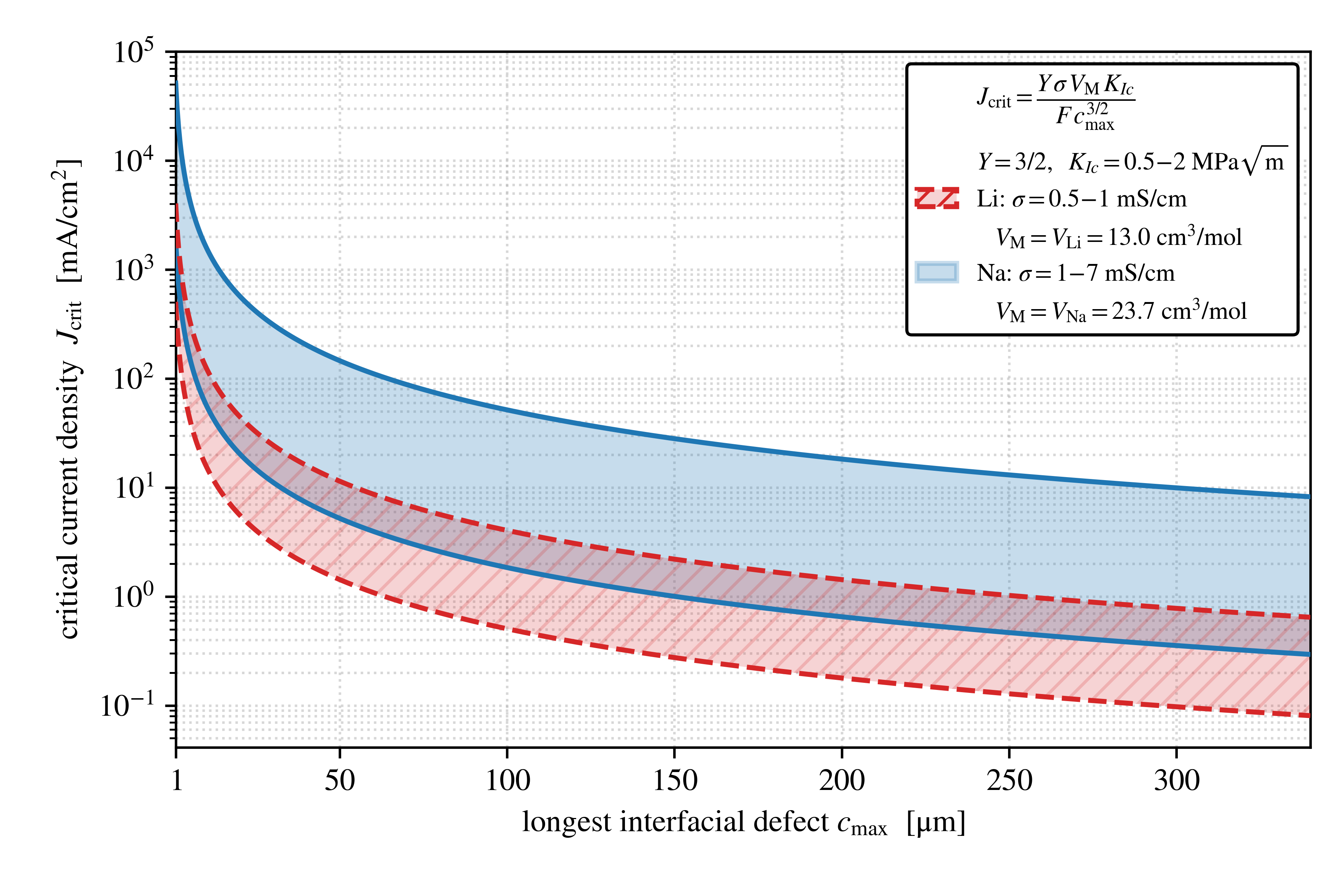}
    \caption{Predicted values of $J_\mathrm{crit}$ depending on the length $c_\mathrm{max}$ of the longest, sufficiently thin interfacial defect.
    }
    \label{figure5}
\end{figure}

\textbf{Values of $J_\mathrm{crit}$:} Typical oxide ceramics values of the fracture toughness $K_\mathrm{Ic}\sim0.5-2\,\mathrm{MPa\,\sqrt{m}}$ are assumed. The ionic conductivity of LLZO (the only oxide ceramic lithium-ion conductor which is sufficiently stable against lithium metal) is roughly in the range of $0.5-1\,\mathrm{mS/cm}$ at room temperature \cite{raju2021crystal, wang2024accelerating}. Oxide ceramic sodium ion conductors which are sufficiently stable against sodium metal (such as NaSICON or NaRSiO) are significantly better conducting. Reported values at room temperatures are in the range $1-7\,\mathrm{mS/cm}$ \cite{Ma.2020, schilm2022, liu2025simultaneously}. Assuming the geometry parameter to be $Y\approx3/2$ and constants $V_\mathrm{Li}=13.0\,\mathrm{cm^3/mol}$, $V_\mathrm{Na}=23.7\,\mathrm{cm^3/mol}$ and $F=96\,485\,\mathrm{C/mol}$, values of $J_\mathrm{crit}$ can be estimated from equation \ref{eq: Dendriten Globale kritische Stromdichte} with dependence on $c_\mathrm{max}$. Insertion yields the results depicted in Figure \ref{figure5}.

\textbf{Comparison with experimental values:} Equation \ref{eq: Dendriten Globale kritische Stromdichte} contains no adjustable parameter besides $Y$ and can therefore be tested against the literature by inverting it: every reported pair of ionic conductivity and critical current density implies a defect length $c_\mathrm{max}$. Here, only oxide ceramics and approximately plating (alkali metal reduction) limited critical current density values are considered. A critical current density measured in a symmetric cell by cycling is only a lower bound, because vacancies accumulate at the interface during stripping (alkali metal oxidation) and thus raise the local current density to an unknown value by decreasing the electrochemically active interfacial contact area \cite{Krauskopf2019, spencerjolly2020sodium}.

\begin{table}[H]
\centering
\small
\begin{tabular}{llcccc}
\hline
metal & electrolyte, authors & protocol & $\sigma$ & $J_\mathrm{crit}$ & $c_\mathrm{max}$\\
 & & & [mS/cm] & [mA/cm$^2$] & [$\mathrm{\mu m}$]\\
\hline
Li & LLZO:Al, Dubey et al. \cite{dubey2021building} & cycling & 0.48 & 0.16 & 333 \\
Li & LLZO:Al, Dubey et al. \cite{dubey2021building} & cycling & 0.48 & 0.32 & 209 \\
Li & LLZTO, Zhu et al. \cite{zhu2025impact} & cycling & 0.41 & 0.34 & 181 \\
Li & LLZTO, Zhu et al. \cite{zhu2025impact} & cycling & 0.53 & 0.70 & 133 \\
Li & LLZTO, Zhu et al. \cite{zhu2025impact} & cycling & 0.64 & 0.56 & 174 \\
Na & Na-$\beta''$-alumina, Spencer Jolly et al. \cite{spencerjolly2020sodium} & unidirectional & 2.0 & 3.0 & 182 \\
Na & NaSmSiO, own measurement \cite{Lowack2025} & unidirectional & 1.5 & 0.96 & 321 \\
Na & NaSICON, Ma et al. \cite{ma2022enhancing} & cycling & 5.2 & 14 & 123 \\
Na & NaSICON:NLP, Liu et al. \cite{liu2025simultaneously} & cycling & 7.1 & 22 & 112 \\
\hline
\end{tabular}
\caption{Previously reported room temperature critical current densities and the defect length which equation \ref{eq: Dendriten Globale kritische Stromdichte} requires to reproduce them, using $K_\mathrm{Ic}=1\,\mathrm{MPa\,\sqrt{m}}$ and $Y=3/2$. The conductivities are those reported for the respective electrolyte in the corresponding publication.}
\label{tab: Literaturvergleich}
\end{table}

The critical current densities in Table \ref{tab: Literaturvergleich} span a factor of $138$, the predicted defect lengths only a factor of $3$, remaining between $112$ and $333\,\mathrm{\mu m}$. Formula \ref{eq: Dendriten Globale kritische Stromdichte} calculates a geometric mean $c_\mathrm{max}$ of $196\,\mathrm{\mu m}$ over the five lithium measurements and of $169\,\mathrm{\mu m}$ over the four sodium ones. Since there is no reason why lithium or sodium ceramics should be more difficult to process, similar values between both material classes are expected. The model reproduces that expectation and sodium samples achieve higher critical current densities compared to lithium due to the larger value of $\sigma\,V_\mathrm{M}$.

\textbf{Absolute values:} While the calculated defect lengths of $c_\mathrm{max}\sim200\,\mathrm{\mu m}$ exhibit reasonable scattering between experiments by different groups, their nominal values are likely unrealistically large. Hence the model appears to overestimate $J_\mathrm{crit}$ systematically by roughly one order of magnitude. The neglected interfacial resistance in the dissipation balance (equation \ref{eq: Dendriten Dissipationsfunktional}) might be one explanation. While homogeneous interface resistance can be numerically included \cite{mukherjee2023ingress}, the real resistive bottleneck at interfaces is likely not homogeneous but rather the inhomogeneous constriction resistance \cite{Krauskopf2020, Ortmann.2023, lowack2025sputtered}. Such incomplete contact effectively means that only some fraction $f$ of the geometric area is in sufficient contact for charge transfer. Hence, a defect experiences a local current density $J/f$ instead of the applied $J$, and the measured critical current density is reduced to $f\,J_\mathrm{crit}$. Since any measurable constriction resistance requires low values of $f$ \cite{Holm1967}, and since already nano-scale pores and contaminations are insurmountable obstacles for alkali ions, values of $f\sim 0.1$ at real interfaces are easily justifiable. Together with the neglected microstructure of the polycrystalline electrolyte which acts in the same direction, this effect must be contained in the prefactor $Y$ of equation \ref{eq: Dendriten Globale kritische Stromdichte}.

\textbf{Statistical scattering:} My previous critical current density measurements on NaSmSiO were best described by a three-parameter Weibull-distribution with a Weibull-modulus of $m=1.1$ \cite{Lowack2025}. Fitting the same data to the more common two-parameter Weibull-distribution yields $m=2.5$ at higher fitting error. The mechanical Weibull-modulus is known to be $m_\mathrm{mech}=2k$, assuming the flaws responsible for electrochemical and mechanical failure share the same Pareto-exponent $k$ (equation \ref{eq: Dendrit cdf-c}). According to equation \ref{weibull modul}, this is three times larger than the electrochemical Weibull-modulus. Therefore, a value of $m_\mathrm{mech}=3.3$ or $m_\mathrm{mech}=7.5$ is expected for the same solid electrolyte, depending if a three- or two-parameter Weibull-distribution is used. While detailed tests on the exact material are still pending, Wagner measured values in the range of $m_\mathrm{mech}=3-9$ on very similar compositions in her doctoral thesis \cite{wagner2022natrium}. Hence, roughly the expected range.

\section*{Conclusion}
A quantitative model of dendrite penetration through solid electrolytes was analytically derived by comparing Joule dissipation during the transport of the ionic current through the solid electrolyte with the mechanical dissipation during interfacial metal deposition. By the thermodynamical argument of minimal total dissipation, and by approximating the shape of interfacial defects as ellipsoids, the critical current density was derived as:
\begin{equation}
\label{eq: Dendriten Globale kritische Stromdichte wiederholung}
    J_\mathrm{crit}\approx\frac{Y\,\sigma\,V_\mathrm{M}\,K_\mathrm{Ic}}{F\,c_\mathrm{max}^{3/2}}
\end{equation}
This expression links $J_\mathrm{crit}$ to the length of the longest interfacial defect $c_\mathrm{max}$ (assuming the defect is sufficiently thin to constrict alkali metal backflow) and to material constants (ionic conductivity $\sigma$ and fracture toughness $K_\mathrm{Ic}$ of the solid electrolyte, molar volume $V_\mathrm{M}$ of the alkali metal, the geometry parameter $Y\sim 3/2$ and the Faraday constant $F=96485\,\mathrm{C\,mol^{-1}}$). 

The formula provides clear and intuitive guidelines for designing solid electrolytes resistant to dendrite penetration. Microstructures containing long and thin defects, such as grain-boundary micro-cracks, drive $J_\mathrm{crit}$ to undesirably low values. Notable progress toward suppressing such defects has been reported by Liu et al. \cite{liu2025simultaneously}. Other strategies include an increase of the ionic conductivity $\sigma$ and the fracture toughness $K_\mathrm{Ic}$. Numerically, the model overestimates $J_\mathrm{crit}$ by about one order of magnitude, an expected result due to the dependence of ion conduction on solid electrolyte microstructure. For samples with non-vanishing and homogeneous interfacial resistance $\gg 1\,\mathrm{\Omega\,cm^2}$, the related but mathematically more complicated model by Mukherjee et al. is recommended \cite{mukherjee2023ingress}.

The defect dependence $\propto c_\mathrm{max}^{-3/2}$ rationalizes the weak-link penetration behavior documented in my previous work \cite{Lowack2025} and explains the pronounced scatter of $J_\mathrm{crit}$ between samples of similar preparation. The model postulates this scattering to be Weibull-distributed. The theoretical Weibull-modulus should approximately be one third of the Weibull-modulus found in mechanical strength tests of the solid electrolyte, which is in good agreement with the experimental observation \cite{Lowack2025}. Following these elaborations, the statistical thinking commonly applied to the mechanical properties of ceramics should be expanded to the critical current densities of ceramic solid electrolytes.

\newpage
\section{Supplementary derivations}

\subsection{Gradient and normal derivative of $\Phi_c$}
To derive the electric field around the ellipsoid, the gradient of equation \ref{eq: Stratton} must be calculated:
\begin{equation}\label{Gradient 1}
    -\nabla \Phi_c(\lambda,z) =\mathbf E(\lambda,z)= \nabla\Bigg\{E_0\,z\left[1 - \frac{\mathcal I(\lambda)}{\mathcal I(0)}\right]\Bigg\}
\end{equation}
Applying the product and chain rule, this yields:
\begin{equation}\label{Gradient 2}
    -\nabla \Phi_c(\lambda,z) =E_0\nabla z\left[1 - \frac{\mathcal I(\lambda)}{\mathcal I(0)}\right]+E_0\,z\!\left[1 - \frac{\mathcal I'(\lambda)}{\mathcal I(0)}\right]\nabla \lambda
\end{equation}
Finding the gradient $\nabla z=\mathbf e_z$ is trivial. Calculating $\mathcal I'(\lambda)$ from equation \ref{eq:Ic} is also trivial, as differentiation simply cancels integration. Finding the gradient $\nabla\lambda$ is an implicit problem of the defining equation of the ellipsoids surface:
\begin{equation}
    \frac{x^2}{w^2+\lambda}+\frac{y^2}{d^2+\lambda}+\frac{z^2}{c^2+\lambda}=1
\end{equation}
Differentiating this equation with respect to $x$ yields:
\begin{equation}
    -\frac{x^2}{(w^2+\lambda)^2}\,\frac{\partial\lambda}{\partial x}+\frac{2x}{w^2+\lambda}-\frac{y^2}{(d^2+\lambda)^2}\,\frac{\partial\lambda}{\partial x}-\frac{z^2}{(c^2+\lambda)^2}\,\frac{\partial\lambda}{\partial x}=0
\end{equation}
Rearranging yields $\frac{\partial\lambda}{\partial x}$. Likewise, differentiation with respect to $y$ and $z$ yields $\frac{\partial\lambda}{\partial y}$ and $\frac{\partial\lambda}{\partial z}$. When defining (for readability) the shorthand
\begin{equation}\label{eq: T}
    T(\lambda)=\frac{x^2}{(w^2+\lambda)^2}+\frac{y^2}{(d^2+\lambda)^2}+\frac{z^2}{(c^2+\lambda)^2},
\end{equation}
this yields:
\begin{equation}
    \nabla\lambda=\frac{2}{T(\lambda)}\bigg(\frac{x}{w^2+\lambda}\,\mathbf e_x+\frac{y}{d^2+\lambda}\,\mathbf e_y+\frac{z}{c^2+\lambda}\,\mathbf e_z\bigg)
\end{equation}
Insertion of $\nabla z$, $\nabla \lambda$ and $\mathcal I'(\lambda)$ (equation \ref{eq:Ic}) into equation \ref{Gradient 2} finally yields:
\begin{equation}\label{Gradient 3}
    -\nabla \Phi_c(\lambda,z) =E_0\left[1 - \frac{\mathcal I(\lambda)}{\mathcal I(0)}\right]\mathbf e_z
    +\frac{E_0\,z\,w\,d\,c}{\mathcal I(0)\,T(\lambda)(c^2+\lambda)\,R(\lambda)}\bigg(\frac{x}{w^2+\lambda}\,\mathbf e_x+\frac{y}{d^2+\lambda}\,\mathbf e_y+\frac{z}{c^2+\lambda}\,\mathbf e_z\bigg)
\end{equation}
By definition, $\lambda=0$ defines the surface $\Gamma_c\cup\Gamma_c'$ of the ellipsoid. Here, and thus on the defect's surface $\Gamma_c$, the gradient should become purely normal. Insertion of $\lambda=0$ and $R(0)$ (equation \ref{eq: R(s)}) into equation \ref{Gradient 3} yields indeed:
\begin{equation}\label{eq: normal derivative 1}
    -\nabla\Phi_c\big|_{\Gamma_c}=\frac{E_0\,z}{\mathcal I(0)\,c^2\,T(0)}\bigg(\frac{x}{w^2}\,\mathbf e_x+\frac{y}{d^2}\,\mathbf e_y+\frac{z}{c^2}\,\mathbf e_z\bigg)=\frac{E_0\,z}{\mathcal I(0)\,c^2\,\sqrt{T(0)}}\,\mathbf n
\end{equation}

\subsection{Derivation of $\Delta \dot Q_\mathrm{max}$}\label{Herleitung Joule Delta}
The aim is to solve the integral of equation \ref{eq: Dendriten maximale Dissipationseinspahrung} to obtain equation \ref{eq: Delta Joule Dissipation}. To ease the derivation, rewrite equation \ref{eq: Dendriten maximale Dissipationseinspahrung} as:
\begin{equation}\label{eq: Q^*}
        \Delta \dot Q^*:=\Delta \dot{Q}_{\mathrm{max}}(J)-\sigma\, E_0^2\,V_c=\sigma\,\int_{\Omega_c} (|\nabla\Phi_c|^2- E_0^2)\,\mathrm dV
\end{equation}
\subsubsection{Volume integration to surface integration via Gauss's theorem}
Begin by rewriting the potential $\Phi_c$ as:
\begin{equation}
\label{eq: Stratton potential rearanged}
    \Phi_c=\Phi_0+\Phi^* 
\end{equation}
$\Phi_0$ is the limit of the electrical potential far away from the defect and thus the potential for the non-localized case of equation \ref{eq: Dendriten Feld trivial}. $\Phi^*$ is the potential perturbation in the vicinity of the defect. Consequently, $\Phi^*$ must decay to zero at infinity and satisfy the Laplace equation, i.e. $\nabla^2\Phi^*=0$. Insertion into equation \ref{eq: Dendriten maximale Dissipationseinspahrung} yields:
\begin{equation}
\label{eq: Dendriten maximale Dissipationseinspahrung_proof}
    \Delta \dot Q^*=\sigma\,\int_{\Omega_c} (|\nabla\Phi_0+\nabla\Phi^*|^2-E_0^2)\,\mathrm dV=\sigma\,\int_{\Omega_c} (2\nabla\Phi_0\,\nabla\Phi^*+|\nabla\Phi^*|^2)\,\mathrm dV
\end{equation}
For simplicity, integrate the mixed and the quadratic term independently.

\textbf{Mixed term:} Using the product rule and $\nabla^2\Phi^*=0$ in ${\Omega_c}$, the integration of the mixed term yields:
\begin{equation}
    \int_{\Omega_c}2\nabla\Phi_0\,\nabla\Phi^*\mathrm dV=2\int_{\Omega_c}\big[\nabla\cdot(\Phi_0\,\nabla\Phi^*)-\Phi_0\,\underbrace{\nabla^2\Phi^*}_{=0}\big]\,\mathrm dV=2\int_{\Omega_c}\nabla\cdot(\Phi_0\,\nabla\Phi^*)\,\mathrm dV
\end{equation}
This expression can be rewritten using Gauss's theorem using the outward pointing unit normal vector $\mathbf n$ of the surface $\partial{\Omega_c}$ of ${\Omega_c}$:
\begin{equation}
\label{eq: mixed Gauss}
    2\int_{\Omega_c}\nabla\cdot(\Phi_0\,\nabla\Phi^*)\,\mathrm dV=2\oint_{\partial{\Omega_c}} \Phi_0\,\nabla\Phi^*\cdot\mathbf{n}\,\mathrm{d}A=2\oint_{\partial{\Omega_c}} \Phi_0\,\frac{\partial\Phi^*}{\partial n}\,\mathrm{d}A
\end{equation}
Following the illustration in Figure \ref{figure4}, $\partial{\Omega_c}$ can now be separated into the ellipsoids surface $\Gamma_c\cup\Gamma'_c$ and a spheroid surface $\partial\Omega_{c,\,\infty}$ of infinite radius, hence:
\begin{equation}
\label{eq: mixed Gauss separation}
    2\oint_{\partial{\Omega_c}} \Phi_0\,\frac{\partial\Phi^*}{\partial n}\,\mathrm{d}A=2\oint_{\Gamma_c\cup\Gamma'_c} \Phi_0\,\frac{\partial\Phi^*}{\partial n}\,\mathrm{d}A+2\oint_{\partial\Omega_{c,\,\infty}} \Phi_0\,\frac{\partial\Phi^*}{\partial n}\,\mathrm{d}A
\end{equation}
The integration over $\partial\Omega_{c,\,\infty}$ vanishes, since $\Phi^*\sim r^{-2}$ far away from the defect (proof by multipole expansion) and thus $\frac{\partial\Phi^*}{\partial n}\sim r^{-3}$. Therefore, the mixed term of equation \ref{eq: Dendriten maximale Dissipationseinspahrung_proof} reduces to an integral over the ellipsoids surface:
\begin{equation}
\label{eq: mixed term}
    \int_{\Omega_c}2\nabla\Phi_0\,\nabla\Phi^*\mathrm dV=2\oint_{\Gamma_c\cup\Gamma'_c} \Phi_0\,\frac{\partial\Phi^*}{\partial n}\,\mathrm{d}A
\end{equation}

\textbf{Quadratic term:} To calculate the quadratic term of equation \ref{eq: Dendriten maximale Dissipationseinspahrung_proof}, consider $\nabla^2\Phi^*=0$ and use the product rule once again:
\begin{equation}
    \int_{\Omega_c} |\nabla\Phi^*|^2\,\mathrm dV=\int_{\Omega_c} \big[\nabla\cdot(\Phi^*\,\nabla\Phi^*)-\Phi^*\,\underbrace{\nabla^2\Phi^*}_{=0}\big]\,\mathrm dV=\int_{\Omega_c}\nabla\cdot(\Phi^*\,\nabla\Phi^*)\,\mathrm dV
\end{equation}
Analogue to equation \ref{eq: mixed Gauss} and \ref{eq: mixed Gauss separation}, Gauss's theorem is again applied and surface integration over $\partial\Omega_{c,\,\infty}$ vanishes:
\begin{equation}
\label{eq: quadratic term}
    \int_{\Omega_c} |\nabla\Phi^*|^2\,\mathrm dV=\oint_{\Gamma_c\cup\Gamma'_c} \Phi^*\,\frac{\partial\Phi^*}{\partial n}\,\mathrm{d}A
\end{equation}

\textbf{Combining both terms:} Now equation \ref{eq: mixed term} and \ref{eq: quadratic term} are inserted into equation \ref{eq: Dendriten maximale Dissipationseinspahrung_proof}, yielding:
\begin{equation}
\label{eq: Dendriten maximale Dissipationseinspahrung_proof_2}
    \Delta \dot Q^*=2\,\sigma\oint_{\Gamma_c\cup\Gamma'_c} \Phi_0\,\frac{\partial\Phi^*}{\partial n}\,\mathrm{d}A+\sigma\oint_{\Gamma_c\cup\Gamma'_c} \Phi^*\,\frac{\partial\Phi^*}{\partial n}\,\mathrm{d}A
\end{equation}
Now consider the boundary condition, the surface of the metal filled defect (i.e. the prolate ellipsoid) and the interface between solid electrolyte and electrode are of constant potential. Since the value of this interfacial potential is well defined up to an additive constant, choose arbitrarily $\Phi_c\big|_{\Gamma_c\cup\Gamma'_c}=0$. Hence equation \ref{eq: Stratton potential rearanged} yields:
\begin{equation}
\label{eq: dendrit Randbedingung}
    \Phi^*\big|_{\Gamma_c\cup\Gamma'_c}=-\Phi_0\big|_{\Gamma_c\cup\Gamma'_c}
\end{equation}
Insertion into equation \ref{eq: Dendriten maximale Dissipationseinspahrung_proof_2} simplifies to:
\begin{equation}
\label{eq: Dendrit Delta Q max Derivation}
    \Delta \dot Q^*=\sigma\oint_{\Gamma_c\cup\Gamma'_c} \Phi_0\,\frac{\partial\Phi^*}{\partial n}\,\mathrm{d}A
\end{equation}
Substituting $\Phi^*$ via equation \ref{eq: Stratton potential rearanged} yields:
\begin{equation}\label{eq: Dendrit Delta Q max Derivation 2}
    \Delta \dot Q^*=\sigma\oint_{\Gamma_c\cup\Gamma'_c} \Phi_0\,\frac{\partial\Phi_c}{\partial n}\,\mathrm{d}A-\sigma\oint_{\Gamma_c\cup\Gamma'_c} \Phi_0\,\frac{\partial\Phi_0}{\partial n}\,\mathrm{d}A
\end{equation}

By the ellipsoid equation, the ellipsoid's surface is in cartesian coordinates an implicit function:
\begin{equation}\label{eq: normal surface parametrization}
    f(x,y,z)=\frac{x^2}{w^2}+\frac{y^2}{d^2}+\frac{z^2}{c^2}-1=0
\end{equation}
By definition and when using the definition of $T(0)$ (equation \ref{eq: T}), the normal unit vector on this surface in cartesian coordinates is thus:
\begin{equation}\label{eq: normal}
    \mathbf n=\frac{\nabla f}{|\nabla f|}=\frac{1}{\sqrt{T(0)}}\bigg(\frac{x}{w^2}\mathbf e_x+\frac{y}{d^2}\mathbf e_y+\frac{z}{c^2}\mathbf e_z\bigg)
\end{equation}
Use this result, to realize that:
\begin{equation}\label{eq: normal derivative 2}
    \frac{\partial\Phi_0}{\partial n}\Bigg|_{\Gamma_c\cup\Gamma'_c}=\nabla \Phi_0\cdot\mathbf n\Bigg|_{\Gamma_c\cup\Gamma'_c}=-E_0\,\mathbf e_z\cdot\mathbf n=-\frac{E_0\,z}{c^2\,\sqrt{T(0)}}
\end{equation}
Equation \ref{eq: normal derivative 1} and \ref{eq: normal derivative 2} are now inserted into equation \ref{eq: Dendrit Delta Q max Derivation 2}. Together with $\Phi_0=-E_0\,z$, this yields:
\begin{equation}\label{eq: Dendrit Delta Q max Derivation 3}
    \Delta \dot Q^*=\frac{\sigma\,E_0^2}{c^2}\,\bigg(\frac{1}{\mathcal I(0)}-1\bigg)\oint_{\Gamma_c\cup\Gamma'_c}\frac{z^2}{\sqrt{T(0)}}\,\mathrm{d}A
\end{equation}

\subsubsection{Solving the surface integral}\label{solving surface integral}
\textbf{Reparametrization:} To solve the integral, reparametrize the ellipsoid's surface in angular coordinates:
\begin{equation}\label{eq: reparametrization}
  x = w\sin\theta\cos\varphi,\quad
  y = d\sin\theta\sin\varphi,\quad
  z = c\cos\theta,
  \quad \theta\in[0,\pi],\;\varphi\in[0,2\pi]
\end{equation}
\textbf{Surface element:} Recall that for a surface which is parametrized by $\mathbf{r}(\theta,\varphi)$,
the surface element is given by:
\begin{equation}\label{eq: surface element 1}
    \mathrm{d}A = |\mathbf{r}_\theta\times\mathbf{r}_\varphi|\,\mathrm{d}\theta\,\mathrm{d}\varphi
\end{equation}
The two tangent vectors on the ellipsoid follow from equation \ref{eq: reparametrization} by component-wise differentiation as: 
\begin{equation}
    \mathbf{r}_\theta = \begin{pmatrix} w\cos\theta\cos\varphi \\ d\cos\theta\sin\varphi \\ -c\sin\theta \end{pmatrix}, \qquad
    \mathbf{r}_\varphi = \begin{pmatrix} -w\sin\theta\sin\varphi \\ d\sin\theta\cos\varphi \\ 0 \end{pmatrix}
\end{equation}
By definition, $\mathbf{r}_\theta$ and $\mathbf{r}_\varphi$ are tangent to the surface normal $\nabla f$ (equation \ref{eq: normal surface parametrization} and \ref{eq: normal}). Consequently, $\mathbf{r}_\theta\times\mathbf{r}_\varphi$ — which is also perpendicular to both tangent vectors — must be parallel to $\nabla f$. Hence
\begin{equation}
    \mathbf{r}_\theta\times\mathbf{r}_\varphi = \xi(\theta,\varphi)\,\nabla f
\end{equation}
for some scalar $\xi$. To find $\xi$, calculate the $z$-component of the
cross product directly:
\begin{equation}
    (\mathbf{r}_\theta\times\mathbf{r}_\varphi)_z
    = (r_\theta)_x(r_\varphi)_y - (r_\theta)_y(r_\varphi)_x
        = w\cos\theta\cos\varphi\cdot d\sin\theta\cos\varphi
    + d\cos\theta\sin\varphi\cdot w\sin\theta\sin\varphi
    = wd\sin\theta\cos\theta
\end{equation}
By equation \ref{eq: normal}, the $z$-component of $\nabla f = 2\big(x/w^2,\,y/d^2,\,z/c^2\big)^T$
evaluated on the surface is:
\begin{equation}
    (\nabla f)_z = \frac{2z}{c^2} = \frac{2c\cos\theta}{c^2} = \frac{2\cos\theta}{c}
\end{equation}
Equating $(\mathbf{r}_\theta\times\mathbf{r}_\varphi)_z = \xi(\nabla f)_z$ yields:
\begin{equation}
    w\,d\sin\theta\cos\theta = \xi\,\frac{2\cos\theta}{c}
    \qquad\Longrightarrow\qquad
    \xi = \frac{w\,d\,c\sin\theta}{2}
\end{equation}
Using $|\nabla f| = 2\sqrt{T(0)}$ from equation \ref{eq: normal} this yields:
\begin{equation}
    |\mathbf{r}_\theta\times\mathbf{r}_\varphi|
    = \xi\,|\nabla f|
    = \frac{w\,d\,c\sin\theta}{2}\cdot 2\sqrt{T(0)}
    = w\,d\,c\,\sqrt{T(0)}\,\sin\theta
\end{equation}
Hence, the surface element follows by insertion in equation \ref{eq: surface element 1} as:
\begin{equation}\label{eq: surface element 2}
    \mathrm{d}A = wdc\sqrt{T(0)}\,\sin\theta\,\mathrm{d}\theta\,\mathrm{d}\varphi
\end{equation}

\textbf{Computation:}
The parametrization of $z$ (equation \ref{eq: reparametrization}) and the surface element $\mathrm dA$ (equation \ref{eq: surface element 2}) are inserted into equation \ref{eq: Dendrit Delta Q max Derivation 3}. This yields:
\begin{equation}
      \oint_S \frac{z^2}{\sqrt{T(0)}}\,\mathrm d A
  = w\,d\,c\int_0^{2\pi}\,\mathrm d\varphi\int_0^\pi c^2\cos^2\!\theta\,\sin\theta\,\mathrm d\theta
  = w\,d\,c^3\cdot 2\pi\cdot\frac{2}{3}
  = \frac{4\pi w\,d\,c^3}{3}
\end{equation}
Insertion into equation \ref{eq: Dendrit Delta Q max Derivation 3} gives:
\begin{equation}
    \Delta \dot Q^*=\bigg(\frac{1}{\mathcal I(0)}-1\bigg)\underbrace{\frac{4\,\pi\,w\,d\,c}{3}}_{V_c}\,\sigma\,E_0^2
\end{equation}
Notice, that the volume $V_c$ of the ellipsoid appears! Inserting this expression back into the original equation \ref{eq: Q^*} yields:
\begin{equation}
    \Delta \dot Q_\mathrm{max}=\frac{4\,\pi\,\sigma\,E_0^2\,w\,d\,c}{3\,\mathcal I(0)}
\end{equation}
Division by $2$, since only the dissipation reduction in the lower half-space is of interest (by the method of image charge) derives the desired expression of equation \ref{eq: Dendriten maximale Dissipationseinspahrung}. \hfill$\Box$

\subsection{Derivation of $I_c$}\label{Herleitung I}
The aim is to solve the integral of equation \ref{eq: Dendriten Absolutstrom} to derive equation \ref{eq: Absolutstrom 2}. Into
\begin{equation}
    I_c=-\sigma\int_{\Gamma_c}\nabla\Phi_c\cdot\mathbf n\,\mathrm{d}A,
\end{equation}
equation \ref{Gradient 3} is inserted for $\nabla \Phi_c$, yielding:
\begin{equation}
    I_c=\frac{\sigma\,E_0}{\mathcal I(0)\,c^2}\int_{\Gamma_c}\frac{z}{\sqrt{T(0)}}\,\mathrm{d}A
\end{equation}
The integral is again parametrized in angular coordinates via equation \ref{eq: reparametrization} and \ref{eq: surface element 2}:
\begin{equation}
\label{eq: Dendriten I_c supplementary}
    I_c=\frac{\sigma\,E_0\,w\,d}{\mathcal I(0)}\int_0^{2\pi}\mathrm d\varphi\int_0^{\pi/2}\cos\theta\,\sin\theta\,\mathrm d\theta=-\frac{\pi\,\sigma\,E_0\,w\,d}{\mathcal I(0)}\big[\cos^2(\theta)\big]^{\pi/2}_0=\frac{\pi\,\sigma\,E_0\,w\,d}{\mathcal I(0)}
\end{equation}
This is the desired expression of equation \ref{eq: Absolutstrom 2}.\hfill $\Box$

\section*{Acknowledgment:} This work was carried out as part of a doctoral thesis at TU Dresden. The author gratefully acknowledges financial support from the State of Saxony within the M.ERA-NET 2023 project "Keramisches Anodenmaterial für definierte Natriumabscheidung" (NaCer-Anode, project 11401).

\bibliographystyle{unsrt}
\bibliography{dendrite_papers.bib}

@article{Porz2017,
  author  = {Porz, Lukas and Swamy, Tushar and Sheldon, Brian W. and
             Rettenwander, Daniel and Fr{\"o}mling, Till and Thaman, Henry L. and
             Berendts, Stefan and Uecker, Reinhard and Carter, W. Craig and
             Chiang, Yet-Ming},
  title   = {Mechanism of Lithium Metal Penetration through Inorganic Solid Electrolytes},
  journal = {Advanced Energy Materials},
  year    = {2017},
  volume  = {7},
  number  = {20},
  pages   = {1701003},
  doi     = {10.1002/aenm.201701003},
}

@article{Klinsmann2019,
  author  = {Klinsmann, Markus and Hildebrand, Felix E. and Ganser, Markus and
             McMeeking, Robert M.},
  title   = {Dendritic cracking in solid electrolytes driven by lithium insertion},
  journal = {Journal of Power Sources},
  year    = {2019},
  volume  = {442},
  pages   = {227226},
  doi     = {10.1016/j.jpowsour.2019.227226},
}

@article{Esmizadeh2025,
  author  = {Esmizadeh, S. and Haftbaradaran, H. and Salvadori, A.},
  title   = {Predicting solid electrolyte fracture by stress-mediated dendrite
             penetration in cracks},
  journal = {International Journal of Mechanical Sciences},
  year    = {2025},
  volume  = {290},
  pages   = {110062},
  doi     = {10.1016/j.ijmecsci.2025.110062},
}

@article{Fincher2022,
  author  = {Fincher, Cole D. and Athanasiou, Christos E. and Gilgenbach, Colin and
             Wang, Michael and Sheldon, Brian W. and Carter, W. Craig and
             Chiang, Yet-Ming},
  title   = {Controlling dendrite propagation in solid-state batteries with
             engineered stress},
  journal = {Joule},
  year    = {2022},
  volume  = {6},
  number  = {12},
  pages   = {2794--2809},
  doi     = {10.1016/j.joule.2022.10.011},
}

@article{Ning2023,
  author  = {Ning, Ziyang and Li, Guanchen and Melvin, Dominic L. R. and
             Chen, Yang and Bu, Junfu and Spencer-Jolly, Dominic and
             Liu, Junliang and Hu, Bingkun and Gao, Xiangwen and
             Perera, Johann and Gong, Chen and Pu, Shengda D. and
             Zhang, Shengming and Liu, Boyang and Hartley, Gareth O. and
             Bodey, Andrew J. and Todd, Richard I. and Grant, Patrick S. and
             Armstrong, David E. J. and Marrow, T. James and
             Monroe, Charles W. and Bruce, Peter G.},
  title   = {Dendrite initiation and propagation in lithium metal solid-state batteries},
  journal = {Nature},
  year    = {2023},
  volume  = {618},
  number  = {7964},
  pages   = {287--293},
  doi     = {10.1038/s41586-023-05970-4},
}

@article{Yuan2021,
  author  = {Yuan, Chunhao and Lu, Wenquan and Xu, Jun},
  title   = {Unlocking the Electrochemical--Mechanical Coupling Behaviors of
             Dendrite Growth and Crack Propagation in All-Solid-State Batteries},
  journal = {Advanced Energy Materials},
  year    = {2021},
  volume  = {11},
  number  = {36},
  pages   = {2101807},
  doi     = {10.1002/aenm.202101807},
}

@article{Xue2025,
  author  = {Xue, Dingchuan and Fincher, Cole D. and Fang, Ruyue and
             Sheldon, Brian W. and Chen, Long-Qing and Zhang, Sulin},
  title   = {Dynamic interplay of dendrite growth and cracking in lithium metal
             solid-state batteries},
  journal = {Journal of the Mechanics and Physics of Solids},
  year    = {2025},
  volume  = {202},
  pages   = {106197},
  doi     = {10.1016/j.jmps.2025.106197},
}

@article{Janek2016,
  author  = {Janek, J{\"u}rgen and Zeier, Wolfgang G.},
  title   = {A solid future for battery development},
  journal = {Nature Energy},
  year    = {2016},
  volume  = {1},
  number  = {9},
  pages   = {16141},
  doi     = {10.1038/nenergy.2016.141},
  url     = {https://www.nature.com/articles/nenergy2016141},
}

@article{Janek2023,
  author  = {Janek, J{\"u}rgen and Zeier, Wolfgang G.},
  title   = {Challenges in speeding up solid-state battery development},
  journal = {Nature Energy},
  year    = {2023},
  volume  = {8},
  number  = {3},
  pages   = {230--240},
  doi     = {10.1038/s41560-023-01208-9},
  url     = {https://www.nature.com/articles/s41560-023-01208-9},
}

@article{Randau2020,
  author  = {Randau, Simon and Weber, Dominik A. and K{\"o}tz, Olaf and
             Koerver, Raimund and Braun, Philipp and Weber, Andr{\'e} and
             Ivers-Tiff{\'e}e, Ellen and Adermann, Torben and Kulisch, J{\"o}rn and
             Zeier, Wolfgang G. and Richter, Felix H. and Janek, J{\"u}rgen},
  title   = {Benchmarking the performance of all-solid-state lithium batteries},
  journal = {Nature Energy},
  year    = {2020},
  volume  = {5},
  number  = {3},
  pages   = {259--270},
  doi     = {10.1038/s41560-020-0565-1},
  url     = {https://www.nature.com/articles/s41560-020-0565-1},
}

@article{Krauskopf2020,
  author  = {Krauskopf, Thorben and Mogwitz, Boris and Hartmann, Hannah and
             Singh, Dheeraj Kumar and Zeier, Wolfgang G. and Janek, J{\"u}rgen},
  title   = {The Fast Charge Transfer Kinetics of the Lithium Metal Anode on the
             Garnet-Type Solid Electrolyte {Li$_{6.25}$Al$_{0.25}$La$_3$Zr$_2$O$_{12}$}},
  journal = {Advanced Energy Materials},
  year    = {2020},
  volume  = {10},
  number  = {27},
  pages   = {2000945},
  doi     = {10.1002/aenm.202000945},
  url     = {https://advanced.onlinelibrary.wiley.com/doi/full/10.1002/aenm.202000945},
}

@article{monroe2005,
  author  = {Monroe, Charles and Newman, John},
  title   = {The Impact of Elastic Deformation on Deposition Kinetics at
             Lithium/Polymer Interfaces},
  journal = {Journal of The Electrochemical Society},
  year    = {2005},
  volume  = {152},
  number  = {2},
  pages   = {A396--A404},
  doi     = {10.1149/1.1850854},
}

@article{Athanasiou2024,
  author  = {Athanasiou, Christos E. and Fincher, Cole D. and Gilgenbach, Colin and
             Gao, Huajian and Carter, W. Craig and Chiang, Yet-Ming and
             Sheldon, Brian W.},
  title   = {Operando measurements of dendrite-induced stresses in ceramic
             electrolytes using photoelasticity},
  journal = {Matter},
  year    = {2024},
  volume  = {7},
  number  = {1},
  pages   = {95--106},
  doi     = {10.1016/j.matt.2023.10.014},
}

@article{fincher2026electrochemical,
  author  = {Fincher, Cole D. and Gilgenbach, Colin and Roach, Christian and
             Osmundsen, Rachel and Penn, Aubrey and Thouless, M. D. and
             Carter, W. Craig and Sheldon, Brian W. and LeBeau, James M. and
             Chiang, Yet-Ming},
  title   = {Electrochemical corrosion accompanies dendrite growth in solid
             electrolytes},
  journal = {Nature},
  year    = {2026},
  volume  = {652},
  number  = {8111},
  pages   = {906--911},
  doi     = {10.1038/s41586-026-10279-z},
}

@article{mukherjee2023ingress,
  author  = {Mukherjee, Dipayan and Hao, Shuai and Shearing, Paul R. and
             McMeeking, Robert M. and Fleck, Norman A. and Deshpande, Vikram S.},
  title   = {Ingress of {Li} into Solid Electrolytes: Cracking and Sparsely
             Filled Cracks},
  journal = {Small Structures},
  year    = {2023},
  volume  = {4},
  number  = {9},
  pages   = {2300022},
  doi     = {10.1002/sstr.202300022},
}

@article{zhang2025atomic,
  author  = {Zhang, Bowen and Yuan, Botao and Yan, Xin and Han, Xiao and
             Zhang, Jiawei and Tan, Huifeng and Wang, Changuo and Yan, Pengfei and
             Gao, Huajian and Liu, Yuanpeng},
  title   = {Atomic mechanism of lithium dendrite penetration in solid electrolytes},
  journal = {Nature Communications},
  year    = {2025},
  volume  = {16},
  number  = {1},
  pages   = {1906},
  doi     = {10.1038/s41467-025-57259-x},
}

@article{Lowack2025,
  author  = {Lowack, Ansgar and Nakum, Yogeshbhai and Anton, Rafael and
             Nikolowski, Kristian and Partsch, Mareike and Michaelis, Alexander},
  title   = {Quantifying Sodium Dendrite Formation in
             {Na$_5$SmSi$_4$O$_{12}$} Solid Electrolytes},
  journal = {Batteries \& Supercaps},
  year    = {2025},
  volume  = {8},
  number  = {12},
  pages   = {e202500279},
  issn    = {2566-6223},
  doi     = {10.1002/batt.202500279},
}

@article{schilm2022,
  author  = {Schilm, Jochen and Anton, Rafael and Wagner, D{\"o}rte and
             Huettl, Juliane and Kusnezoff, Mihails and Herrmann, Mathias and
             Kim, Hong Ki and Lee, Chang Woo},
  title   = {Influence of {R} = {Y}, {Gd}, {Sm} on Crystallization and Sodium Ion
             Conductivity of {Na$_5$RSi$_4$O$_{12}$} Phase},
  journal = {Materials},
  year    = {2022},
  volume  = {15},
  number  = {3},
  pages   = {1104},
  doi     = {10.3390/ma15031104},
}

@article{liu2025simultaneously,
  author  = {Liu, Limin and Ma, Qianli and Zhou, Xiaoliang and Ding, Ziming and
             Gr{\"u}ner, Daniel and K{\"u}bel, Christian and Tietz, Frank},
  title   = {Simultaneously improving sodium ionic conductivity and dendrite
             behavior of {NaSICON} ceramics by grain-boundary modification},
  journal = {Journal of Power Sources},
  year    = {2025},
  volume  = {626},
  pages   = {235773},
  doi     = {10.1016/j.jpowsour.2024.235773},
}

@article{shapovalov2017,
  author  = {Shapovalov, V. M.},
  title   = {On the Applicability of the {Ostwald--de Waele} Model in Solving
             Applied Problems},
  journal = {Journal of Engineering Physics and Thermophysics},
  year    = {2017},
  volume  = {90},
  number  = {5},
  pages   = {1213--1218},
  doi     = {10.1007/s10891-017-1676-9},
}

@phdthesis{wagner2022natrium,
  author  = {Wagner, D{\"o}rte},
  title   = {Natrium-ionenleitende Glaskeramiken zur Anwendung in {Na}-Batterien},
  year    = {2022},
  school  = {Technische Universit{\"a}t Dresden},
  address = {Dresden},
  type    = {Dissertation},
}

@book{stratton1941,
  author    = {Stratton, Julius Adams},
  title     = {Electromagnetic Theory},
  series    = {International Series in Pure and Applied Physics},
  publisher = {McGraw-Hill Book Company},
  address   = {New York},
  year      = {1941},
  pages     = {615},
  url       = {https://archive.org/details/electromagnetict031016mbp},
}

@book{Landau1984,
  author    = {Landau, Lev Davidovich and Lifshitz, Evgeny Mikhailovich and
               Pitaevskii, Lev Petrovich},
  title     = {Electrodynamics of Continuous Media},
  series    = {Course of Theoretical Physics},
  volume    = {8},
  edition   = {2},
  publisher = {Pergamon Press},
  address   = {Oxford},
  year      = {1984},
  isbn      = {978-0-08-030275-1},
}

@article{Ma.2020,
 author = {Ma, Qianli and Tietz, Frank},
 year = {2020},
 title     = {Solid-State Electrolyte Materials for Sodium Batteries: Towards Practical Applications},
 pages = {2693--2713},
 volume = {7},
 number = {13},
 issn = {2196-0216},
 journal = {ChemElectroChem},
 doi = {10.1002/celc.202000164}
}

@article{Krauskopf2019,
  author  = {Krauskopf, Thorben and Hartmann, Hannah and Zeier, Wolfgang G. and
             Janek, J{\"u}rgen},
  title   = {Toward a Fundamental Understanding of the Lithium Metal Anode in
             Solid-State Batteries -- An Electrochemo-Mechanical Study on the
             Garnet-Type Solid Electrolyte {Li$_{6.25}$Al$_{0.25}$La$_3$Zr$_2$O$_{12}$}},
  journal = {ACS Applied Materials \& Interfaces},
  year    = {2019},
  volume  = {11},
  number  = {15},
  pages   = {14463--14477},
  doi     = {10.1021/acsami.9b02537},
}

@article{spencerjolly2020sodium,
  author  = {Spencer Jolly, Dominic and Ning, Ziyang and Darnbrough, James E. and
             Kasemchainan, Jitti and Hartley, Gareth O. and Adamson, Paul and
             Armstrong, David E. J. and Marrow, James and Bruce, Peter G.},
  title   = {Sodium/{Na} $\beta''$ Alumina Interface: Effect of Pressure on Voids},
  journal = {ACS Applied Materials \& Interfaces},
  year    = {2020},
  volume  = {12},
  number  = {1},
  pages   = {678--685},
  doi     = {10.1021/acsami.9b17786},
}

@article{Ortmann.2023,
  author  = {Ortmann, Till and Burkhardt, Simon and Eckhardt, Janis Kevin and
             Fuchs, Till and Ding, Ziming and Sann, Joachim and Rohnke, Marcus and
             Ma, Qianli and Tietz, Frank and Fattakhova-Rohlfing, Dina and
             K{\"u}bel, Christian and Guillon, Olivier and Heiliger, Christian and
             Janek, J{\"u}rgen},
  title   = {Kinetics and Pore Formation of the Sodium Metal Anode on
             {NASICON}-Type {Na$_{3.4}$Zr$_2$Si$_{2.4}$P$_{0.6}$O$_{12}$} for
             Sodium Solid-State Batteries},
  journal = {Advanced Energy Materials},
  year    = {2023},
  volume  = {13},
  number  = {5},
  pages   = {2202712},
  doi     = {10.1002/aenm.202202712},
}

@book{Holm1967,
  author    = {Holm, Ragnar},
  title     = {Electric Contacts: Theory and Application},
  edition   = {4th},
  publisher = {Springer},
  address   = {Berlin, Heidelberg},
  year      = {1967},
  doi       = {10.1007/978-3-662-06688-1},
}

@article{lowack2025sputtered,
  title={Sputtered Zero-Excess Electrodes With Metallic Seed Layers for Solid-State Sodium Batteries},
  author={Lowack, Ansgar and Grun, Paula and Anton, Rafael and Auer, Henry and Nikolowski, Kristian and Partsch, Mareike and Kusnezoff, Mihails and Michaelis, Alexander},
  journal={Batteries \& Supercaps},
  volume={8},
  number={5},
  pages={e202400364},
  year={2025},
  publisher={Wiley Online Library}
}

@article{dubey2021building,
  author  = {Dubey, Romain and Sastre, Jordi and Cancellieri, Claudia and
             Okur, Faruk and Forster, Alexander and Pompizii, Lea and
             Priebe, Agnieszka and Romanyuk, Yaroslav E. and
             Jeurgens, Lars P. H. and Kovalenko, Maksym V. and
             Kravchyk, Kostiantyn V.},
  title   = {Building a Better {Li}-Garnet Solid Electrolyte/Metallic {Li}
             Interface with Antimony},
  journal = {Advanced Energy Materials},
  year    = {2021},
  volume  = {11},
  number  = {39},
  pages   = {2102086},
  doi     = {10.1002/aenm.202102086},
}

@article{zhu2025impact,
  author  = {Zhu, Jie and Wu, Yunfan and Zhang, Hongyi and Xie, Xujia and
             Yang, Yong and Peng, Hongyu and Liang, Xiaochun and Qi, Qiongqiong and
             Lin, Weibin and Peng, Dong-Liang and Wang, Laisen and Lin, Jie},
  title   = {Impact of compaction pressure on formation and performance of
             garnet-based solid-state lithium batteries},
  journal = {Energy Materials},
  year    = {2025},
  volume  = {5},
  number  = {4},
  pages   = {500034},
  doi     = {10.20517/energymater.2024.201},
}

@article{ma2022enhancing,
  author  = {Ma, Qianli and Ortmann, Till and Yang, Aikai and Sebold, Doris and
             Burkhardt, Simon and Rohnke, Marcus and Tietz, Frank and
             Fattakhova-Rohlfing, Dina and Janek, J{\"u}rgen and Guillon, Olivier},
  title   = {Enhancing the Dendrite Tolerance of {NaSICON} Electrolytes by
             Suppressing Edge Growth of {Na} Electrode along Ceramic Surface},
  journal = {Advanced Energy Materials},
  year    = {2022},
  volume  = {12},
  number  = {40},
  pages   = {2201680},
  doi     = {10.1002/aenm.202201680},
}

@article{raju2021crystal,
  title={Crystal structure and preparation of Li7La3Zr2O12 (LLZO) solid-state electrolyte and doping impacts on the conductivity: an overview},
  author={Raju, Md Mozammal and Altayran, Fadhilah and Johnson, Michael and Wang, Danling and Zhang, Qifeng},
  journal={Electrochem},
  volume={2},
  number={3},
  pages={390--414},
  year={2021},
  publisher={MDPI}
}

@article{wang2024accelerating,
  title={Accelerating the development of LLZO in solid-state batteries toward commercialization: a comprehensive review},
  author={Wang, Yang and Chen, Zhen and Jiang, Kai and Shen, Zexiang and Passerini, Stefano and Chen, Minghua},
  journal={Small},
  volume={20},
  number={35},
  pages={2402035},
  year={2024},
  publisher={Wiley Online Library}
}

\end{document}